\documentclass[prd,aps,eqsecnum,floatfix,nofootinbib,preprint,tightenlines]{revtex4}

\usepackage{latexsym}
\usepackage{amsmath}
\usepackage{graphicx}
\def\floatcaption#1#2{ \caption{#2 \label{#1}} }

\def\bibi{\bibitem}

\let\inodot=\i

\def\a{\alpha}

\def\c{\chi}
\def\d{\delta}
\def\e{\epsilon}                % Also, \varepsilon
\def\g{\gamma}

\def\i{\iota}

\def\k{\kappa}
\def\l{\lambda}
\def\m{\mu}
\def\n{\nu}

\def\p{\pi}                     % Also, \varpi
\def\r{\rho}                    %       \varrho
\def\s{\sigma}                  %       \varsigma
\def\t{\tau}

\def\D{\Delta}

\def\G{\Gamma}

\def\P{\Pi}

\def\cc{{\cal C}}

\def\cl{{\cal L}}

\def\co{{\cal O}}

\def\cbo{{\,\raise-.15ex\Sc [\,}}                       % curly "
\def\svev#1{\left\langle #1\right\rangle}       % variable < >
\def\ddt#1{{\buildrel {\hbox{\LARGE .\kern-2pt.}} \over {#1}}}% double dot-over
\def\ie{\mbox{\it i.e.}}
\def\eg{\mbox{\it e.g.}}
\def\etc{\mbox{\it etc.}}

\def\tr{{\rm tr}\,}

\def\half{{1\over 2}}

\def\ttl#1{{\it #1}}
\long \def \blockcomment #1\endcomment{}

\def\lmssq{m^2_\s  L^2}
\def\lmpsq{m^2_\p L^2}

\def\hx{\hat{x}}

\def\psibar{{\overline{\psi}}}

\begin{document}

\begin{center}
{\large\bf On the application of Effective Field Theory
to \\ finite-volume effects in \begin{boldmath}$a_\m^{\rm HVP}$\end{boldmath} }\\
\vspace{3ex}
{Christopher~Aubin,$^a$ Thomas~Blum,$^b$
Maarten~Golterman,$^c$
Santiago~Peris$^d$
\\[0.1cm]
{\it
\null$^a$Department of Physics and Engineering Physics,\\ Fordham University, Bronx,
New York, NY 10458, USA\\
\null$^b$Physics Department,\\
University of Connecticut, Storrs, CT 06269, USA\\
\null$^c$Department of Physics and Astronomy, San Francisco State University,\\ San Francisco, CA 94132, USA\\
\null$^d$Department of Physics and BIST, Universitat Aut\`onoma de Barcelona,\\
E-08193 Bellaterra, Barcelona, Spain}}
\\[6mm]
{ABSTRACT}
\\[2mm]
\end{center}
\begin{quotation}
One of the more important systematic effects affecting lattice computations of the
hadronic vacuum polarization contribution to the
anomalous magnetic moment of the muon, $a_\m^{\rm HVP}$, is the distortion due to a finite
spatial volume.   In order to reach sub-percent precision, these effects
need to be reliably estimated and corrected for, and one of the methods that has been
employed
for doing this is finite-volume chiral perturbation theory.   In this paper, we argue that
finite-volume corrections to $a_\m^{\rm HVP}$ can, in principle, be calculated at any given order
in chiral perturbation theory.   More precisely, once all low-energy constants needed to
define the Effective Field Theory representation of $a_\m^{\rm HVP}$  in infinite
volume are known to a given order, also the finite-volume corrections can be 
predicted to that order in the chiral expansion.   
\end{quotation}

\section{\label{intro} Introduction}
Recent years have seen renewed efforts to obtain a more reliable and more precise
Standard-Model estimate of the muon anomalous magnetic
moment.   The discrepancy between the best Standard-Model estimates and the experimental
value from the Brookhaven experiment \cite{BNL} has not only persisted, but also has
become more acute, as many of the systematic errors associated with the Standard-Model
estimate have become more controlled.   According to a recent review of the Standard-Model
calculation \cite{WP} the best average Standard-Model value is estimated to be  $3.7\s$ smaller than the value reported in Ref.~\cite{BNL}.

The renewed efforts to improve the Standard-Model estimate are driven by new experimental
programs at both Fermilab \cite{Fermilab} and J-PARC \cite{JPARC}, which aim to improve
on the precision of the measurement of the magnetic moment, in the case of Fermilab, by a
factor four.   It is thus important to improve the precision of the Standard-Model estimate
to a level commensurate with the experimental goal.

The bulk of the error of the Standard-Model estimate originates from the hadronic
contributions to the anomalous magnetic moment, which cannot be computed in perturbation
theory.   The hadronic contribution consists of two parts: the hadronic vacuum polarization (HVP)
contribution and the hadronic light-by-light contribution; in this paper, the focus will be the
HVP contribution.   The most precise estimate has been based on the dispersive, data-driven
approach, but more recently lattice QCD computations have started to become more precise,
and are expected to become competitive in the near future.   It is thus important to gain a
thorough understanding of the various systematic errors afflicting lattice computations of the
HVP contribution to the muon anomalous magnetic moment, $a_\m^{\rm HVP}$.

Lattice computations of the HVP are necessarily done in a
finite physical spatial volume, typically a cubic volume with periodic boundary
conditions.\footnote{The euclidean time extent of the lattice is usually significantly
larger than the linear spatial dimension.   We will not consider effects of the finite
extent of the lattice in euclidean time in this paper.}
 As $a_\m^{\rm HVP}$ is dominated by momenta at a scale set by the muon mass, which is roughly equal to the pion mass,
it turns out that finite-volume (FV) effects constitute
one of the more important systematic errors in state-of-the-art lattice computations of the
HVP.   These effects are large enough that lattice results for
$a_\m^{\rm HVP}$ need to be corrected, and it is thus important to compute this
correction, as well as the systematic errors associated with such a correction, reliably.
Various methods have been used to provide reliable estimates:  an adaptation
\cite{Mainz}
of the Gounaris--Sakurai model \cite{GS} for the low-momentum HVP
to finite volume using the method of Ref.~\cite{LL}; chiral perturbation theory (ChPT) \cite{us1,BR,us2,BMW}
in finite volume \cite{GL}; a model for low-energy pions including the $\r$ and $\omega$
resonances \cite{HPQCD}; a systematic estimate of the leading FV effects in terms of the forward Compton amplitude of the pion \cite{HP1,HP2}, based on the methods of Ref.~\cite{ML}; and, finally, by
varying the lattice volume directly in the numerical computations of $a_\m^{\rm HVP}$ \cite{Shetal,BMW}.

Because the relevant scale is so low, the proper systematic Effective Field Theory (EFT) to analyze these FV effects is Chiral Perturbation Theory (ChPT).
In Ref.~\cite{us2}, we computed FV corrections to
NNLO, \ie, to two loops in ChPT.\footnote{A small mistake was corrected in Ref.~\cite{BMW}; for the computation
of Ref.~\cite{us2} the numerical effect of this mistake is negligibly small.}
Since the lowest-order pionic contribution to $a_\m^{\rm HVP}$ already involves a pion loop, we will
follow the convention of referring to the lowest order contribution as NLO, the contribution
that involves two loops in ChPT as NNLO, \etc

Our motivation here
is not to push this to higher orders, but to consider whether, as a matter of principle, FV
corrections for $a_\m^{\rm HVP}$ can be computed to arbitrary orders in ChPT.
Even if orders beyond NNLO may never be pursued in practice, it is important to
establish that ChPT allows, in principle, a systematic approach to FV effects which is
well-defined at each order in the chiral expansion. Our main motivation is to illustrate how the properties of an EFT guarantee this to happen through
simple examples.

The HVP contribution to the muon anomalous
magnetic moment to lowest order in $\a$ is given by \cite{TB} (see also Ref.~\cite{LPR})
\begin{equation}
\label{muonan}
a_\m^{\rm HVP}=4\a^2\int_0^\infty dq^2 f(q^2)\hat\P(q^2)\ ,
\end{equation}
where
\begin{equation}
\label{subvacpol}
\hat\P(q^2)=\P(0)-\P(q^2)\ ,
\end{equation}
is the subtracted HVP, obtained from
\begin{equation}
\label{jj}
\left(q^2\d_{\m\n}-q_\m q_\n\right)\P(q^2)=\int d^4x\, e^{iqx}\svev{j_\m^{\rm EM}(x)j_\n^{\rm EM}(0)}\ ,
\end{equation}
with $j_\m^{\rm EM}(x)$ the hadronic part of the electromagnetic current and $\a$ the fine-structure
constant.   Here the momentum
$q$ is euclidean, and throughout this paper, we will work in euclidean space.  The weight
$f(q^2)$ in Eq.~(\ref{muonan}) depends on the muon mass $m_\m$, and is given by
\begin{eqnarray}
\label{f}
f(q^2)&=&\frac{m_\m^2 q^2 Z^3(q^2)(1-q^2 Z(q^2))}{1+m_\m^2 q^2 Z^2(q^2)}\ ,\\
Z(q^2)&=&\frac{\sqrt{1+4m_\m^2/q^2}-1}{2m_\m^2}\ .\nonumber
\end{eqnarray}
The integral in Eq.~(\ref{muonan}) is finite in
QCD, as can be seen from the operator product expansion of $j_\m^{\rm EM}(x)j_\n^{\rm EM}(0)$, which governs the behavior of $\hat\P(q^2)$
at large $q^2$.   However, ChPT, which is designed to parametrize the small-$q^2$
behavior of $\hat\P(q^2)$, does not get the large-$q^2$ behavior right, and therefore, when we
insert a ChPT representation of $\hat\P(q^2)$ into Eq.~(\ref{muonan}),  the integral
over $q^2$ does not converge beyond some order.   In fact, at $k$-loop order in ChPT,
\begin{equation}
\label{ChPTdiv}
\hat\P(q^2)\sim (q^2)^{k-1}\ ,\quad k=1,\ 2,\ \dots\ ,
\end{equation}
modulo logarithmic corrections.
Because  $f(q^2)\sim m_\m^4/q^6$ for large $q^2$, this
means that the ChPT result for $a_\m^{\rm HVP}$ is finite only up to NNLO, \ie, $k=2$, while at N$^3$LO and higher the integral in Eq.~(\ref{muonan}) is UV divergent,
and new counter terms need to be introduced to render the result finite.   Such counter terms
introduce new low-energy constants (LECs), and thus ChPT cannot be used to estimate $a_\m^{\rm HVP}$
quantitatively beyond two-loop order, unless it would be possible to estimate the value of the
(renormalized) LECs from some other physical processes.   

A key point is that these new counter
terms arise because of the integral over $q^2$ in Eq.~(\ref{muonan}), and they are thus not part
of the ChPT lagrangian used to calculate $\P(q^2)$ to any given order.   They will necessarily
be constructed not only from the pion fields of ChPT but also from the muon and photon fields.
As the photon is massless, and the muon is lighter than a pion, this
raises the additional question whether it is consistent to consider only
FV effects associated with pions, while muons and photons are kept in an
infinite volume.

In Ref.~\cite{us2} we claimed that, nevertheless, pionic FV corrections can be computed in
ChPT to all orders, based on the notion that FV effects refer to the IR behavior
of the theory, while counter terms fix the UV behavior.  As we will see,
the way this separation of scales works is subtle, and the claim is perhaps not obvious.   Also
in a finite volume $\hat\P(q^2)$ takes the form of an expansion in powers of $q^2$ in ChPT,
and one might fear that thus also the FV part of $a_\m^{\rm HVP}$, defined as in Eq.~(\ref{muonan}) with
$\hat\P(q^2)$ replaced by its FV part, diverges beyond NNLO in ChPT.\footnote{We
will argue that, in fact, the FV part of Eq.~(\ref{muonan}) starts diverging at N$^4$LO.}
This concern, that the computation of higher-order FV corrections to $a_\m^{\rm HVP}$
in ChPT might break down, was raised in Ref.~\cite{HP2}.

In this paper, we will argue that, nevertheless, FV corrections to $a_\m^{\rm HVP}$
can be computed at any order in ChPT.   More precisely, our claim is that, once all
counter terms needed to make $a_\m^{\rm HVP}$ finite in a ChPT calculation in
{\it infinite} volume have been introduced, also the finite-volume corrections to
$a_\m^{\rm HVP}$ will be UV finite, with no need to introduce any further counter terms.
In addition, this also holds when only the hadronic part of Eq.~(\ref{muonan}) is considered
in a finite volume, while the muons and photons (the QED part of Eq.~(\ref{muonan})) are
kept in infinite volume.   This is actually the situation encountered in lattice QCD
computations of $a_\m^{\rm HVP}$, in which only $\hat\P(q^2)$ is calculated on the
lattice, and thus in a finite volume.

It is beyond the scope of this paper to give an all-order proof, or even to carry out
explicit ChPT calculations beyond NNLO.   Instead, we will present general
arguments supporting our claim, and discuss the form of the new counter terms
introduced to absorb UV divergences which arise in a ChPT calculation of the $q^2$ integral in Eq.~(\ref{muonan}) in more detail.   This is done in Sec.~\ref{CTFV}.   Then, in
Sec.~\ref{toymodel}, we invent a toy model that allows us to demonstrate how our
claim works already at two-loop order, thus illustrating the mechanism underlying our claim
in a simple example.    We end with our conclusions.

There are two appendices.   In the first appendix, we discuss the relation between
counter terms in the momentum representation for $a_\m^{\rm HVP}$, Eq.~(\ref{muonan}),
which is exclusively used in the main text,
and the ``time-momentum'' representation \cite{BM} often employed in lattice
computations of $a_\m^{\rm HVP}$.   The other appendix contains a number of
technical details needed for Secs.~\ref{CTFV} and \ref{toymodel}.

\section{\label{CTFV} Counter terms for $a_\m^{\rm HVP}$}

In Sec.~\ref{qual}, we begin with a qualitative discussion of the problem, based on a diagrammatic
picture, explaining the necessary introduction of counter terms not present in the low-energy
pion effective theory.   Then, in Sec.~\ref{PauliCT}, we discuss the explicit form of these
counter terms in more detail, and in Sec.~\ref{example}  we present
illustrative examples of the role of these counter terms.   We will have a first look at
FV effects in Sec.~\ref{exfinvol}, to argue that the interplay between UV divergences and
FV effects only becomes non-trivial at N$^4$LO.
While FV effects
will also be qualitatively discussed in Sec.~\ref{qual} below, most of the finite-volume
discussion will be postponed to Sec.~\ref{toymodel}; the bulk of this section will
concentrate on the counter-term structure for $a_\m^{\rm HVP}$ in infinite volume.

\subsection{\label{qual} Qualitative discussion}
In Fig.~\ref{Sample}, diagram N0 depicts the standard diagrammatic picture of the HVP contribution to the muon anomalous magnetic moment.
According to  Refs.~\cite{HP1,HP2}, the HVP contribution in Eq.~(\ref{muonan}) may, in more detail, be thought of in terms of a forward pion Compton scattering subdiagram.   This is schematically depicted in the panel N1 of
Fig.~\ref{Sample} where the internal Compton amplitude is obtained by cutting open the pion loop. In this diagram the fat line depicts any number of pions as well as any other heavier hadrons, such as \eg\ the $\r$, $\omega$, \etc\ resonance contributions or a proton loop, which are denoted in the diagram by $H$. In this picture, ChPT is the result of ``integrating out" all heavier states denoted by $H$, giving rise to the low-energy Effective Field Theory (EFT) which is ChPT.\footnote{While the diagrams show only one pion loop explicitly, there can be more than one pion loop.}  Although it is not known
how to make this quantitative in the real world, it presents a clear picture which will prove helpful in what follows.

Diagram N1$^{\rm EFT}$ schematically depicts the contribution of the low-energy degrees of freedom, \ie, the pions, while diagrams N2-N5 originate from  integrating out all heavier hadronic
contributions, collectively denoted as $H$ in diagram N1. As the $H$ states are considered infinitely heavier than the pions in the setting of the EFT, all $H$ propagators get shrunk to a point, as seen, \eg,  in diagram N2.  Only pion
loops remain, with one of them schematically indicated in N2.   The sum of N1$^{\rm EFT}$ and N2
has a different UV behavior than diagram N1, which must be compensated by the necessary low-energy constants which play the role of counter terms to subtract the UV divergences which are
produced by the ``incorrect'' large-momentum behavior of the sum of N1$^{\rm EFT}$ and N2.
These counter-term contributions correspond to the diagrams denoted by N3-N5.  Both
the loop containing the muon line as well as pion loops become more divergent as a 
consequence of contracting
heavy-hadron propagators to a point.  Diagram N3 renormalizes the UV sub-divergence from the
muon-photon loop in N1$^{\rm EFT}$ and N2,\footnote{While the muon-photon(-pion) loop in N1$^{EFT}$ and N2 looks convergent, at sufficiently high order in ChPT
there will be derivatives at the photon-pion vertex making this loop divergent.} while N4 renormalizes the UV sub-divergence from
the pion loop in N1$^{\rm EFT}$ and N2. Diagram N5 is needed to renormalize the product of these
divergences.   Each of these new ``vertices" (denoted by solid squares  in diagrams N2-N5) corresponds to a set of new higher-dimension operators suppressed by the right power of the scale  $M_H$ characterizing the heavy states $H$. In this sense, ChPT is nothing but an expansion in
inverse powers of $M_H$.  As is clear from these diagrams, one should expect not only an LEC associated with the HVP subdiagram (N4) but also a mixed one involving pions, muons and photons (N3), as well as a counter term without pions (N5). Integrating out $H$, therefore, yields an EFT which is not just ChPT but ChPT enlarged by the presence of muons and photons; in other words, what we need is the EFT for low-energy QCD plus QED.  Diagram
N4 corresponds to a counter term in standard ChPT, but both N3 and N5 correspond
to Pauli-like counter terms coupling to the muon and the photon, with N3 coupling to two pions as well.

\begin{figure}[!ht]
\begin{center}
\includegraphics[width=6in,origin=c]{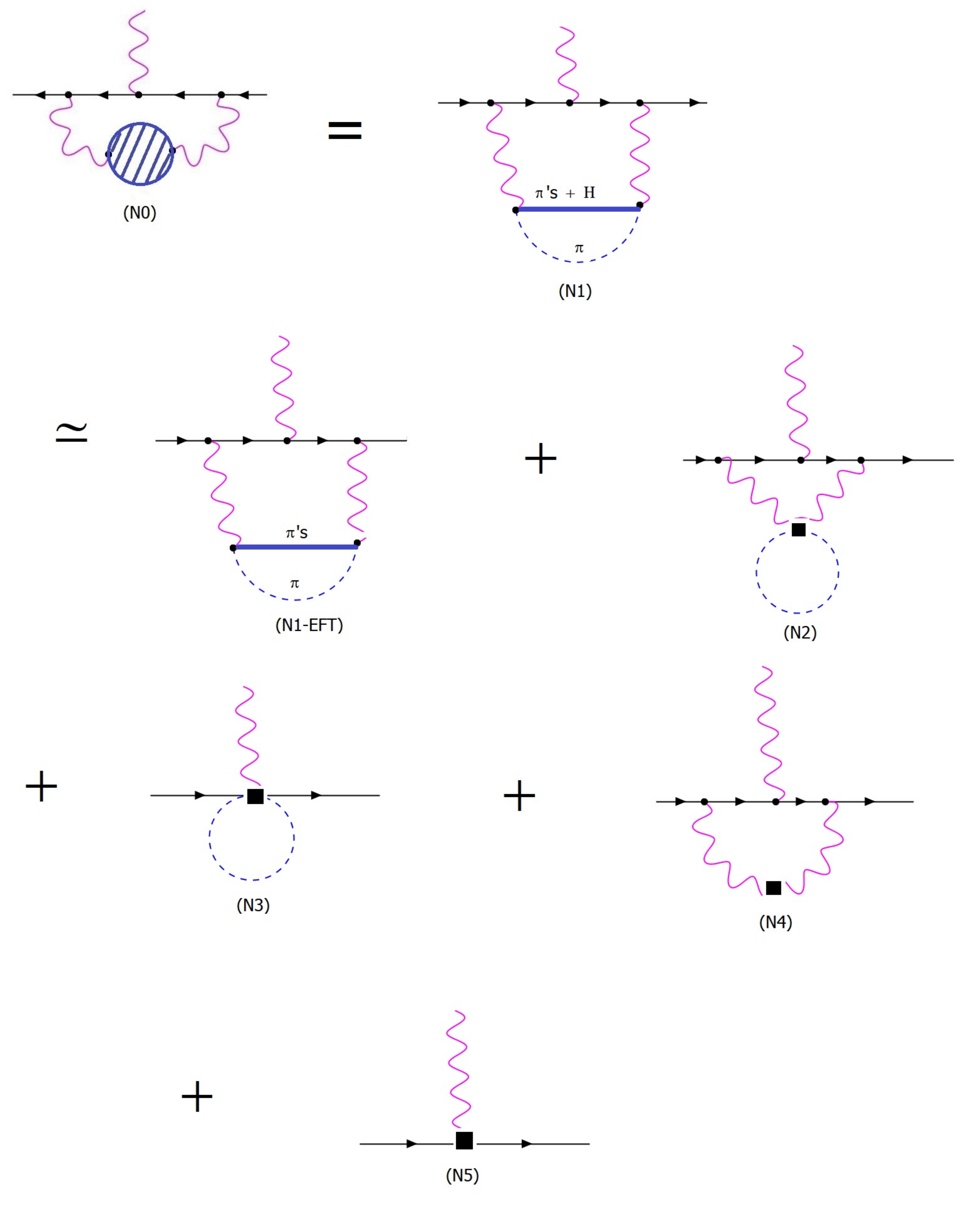}
\floatcaption{Sample}{$a_\m^{\rm HVP}$ in terms of hadronic Feynman diagrams.}
\end{center}
\end{figure}

So far, our discussion has been restricted to infinite volume, and we now turn to the
implications for the EFT in a finite volume.

While the analysis in this paper will be carried out in momentum space, it is useful to
think about the origin of FV effects in position space.
Pion induced FV corrections to the HVP are obtained by having (at least) one pion wrap
(at least once) around the ``periodic world,'' thus producing a factor $e^{-nm_\p L}$, where
$L$ is the linear size of the periodic spatial volume \cite{GL}.   Here $n$ is a number relating to how
many times in which directions the pion wraps around the world: if the pion wraps once in one direction, $n=1$; if it wraps once in two different directions, $n=\sqrt{2}$, \etc\    Since a pion wrapping at least once around the world travels a long distance, it will not contribute to
the degree of divergence of the diagram.
However, the remainder of the diagram may still
be UV divergent, and one may thus end up with a FV correction multiplied by a UV divergence.
This appears to be in conflict with the claim that FV corrections can be systematically computed in an EFT framework.
{As we will see explicitly in Sec.~\ref{toymodel}, in 
momentum space, forcing a pion to wrap around the periodic world corresponds to 
cutting the corresponding pion loop, \ie, putting that pion on shell.

Within ChPT, there will be a counter term corresponding to the divergent subdiagram.   At
orders beyond the order at which the counter term first appears, it will contribute to loop diagrams as well,
and again cutting pion lines on these loops will also lead to FV corrections of the same form,
multiplied by the LEC associated with the counter term.   Since this LEC is UV divergent,
this will renormalize the coefficient of the FV correction, and the complete FV correction
will turn out to be finite, thus exhibiting the expected separation of UV and IR effects in the EFT.
FV effects in $\hat{\P}(q^2)$  are thus predicted in terms of the LECs of  ChPT.  An explicit example of this appears in Ref.~\cite{us2} at NNLO, where the $\co(p^4)$ counter term
$\ell_6$ appears inside a one-loop diagram.   The pion line in this diagram can
wrap around the world, and thus the diagram contributes to FV effects, yielding a FV contribution
proportional to $\ell_6$.   But, FV effects also appear in the two-loop diagrams that appear at
this order.   Taking the pion on one of the loops around the world still leaves a one-loop subdivergence.   This one-loop subdivergence is renormalized by $\ell_6$, and
the sum of the FV effects coming from the two-loop diagrams and the one-loop
diagrams containing $\ell_6$ is UV finite.   The vertex corresponding to the counter term with
LEC $\ell_6$ is an example of a diagram of type N4 in Fig.~\ref{Sample}.

In the case of $a_\m^{\rm HVP}$, there appear also counter terms corresponding to the black squares in
diagrams N3 and N5.  The reason for this is as follows. $a_\m^{\rm HVP}$ is proportional to the projection
on the Dirac structure associated with the muon form factor $F_2(q^2)$ at low momentum of a correlation function in QCD coupled
to muons and photons.   This enlarged theory is described at low energy by a theory of
pions coupled to muons and photons (which, both being lighter than pions, have to be kept
as explicit degrees of freedom in the EFT).   In this extended EFT, new counter terms
can appear which contain muon and photon fields  combining into a Pauli-like
operator.   While in the renormalizable UV-complete theory consisting of QCD plus QED such a
counter term cannot (and does not) appear since it has at least dimension five, in the
EFT there is no restriction on the dimension of possible counter terms, because an
arbitrary inverse power of the heavy scale $M_H$ that has been integrated out can appear
multiplying these counter terms.  Pauli-like
counter terms are thus expected to appear.   In the next subsection, we will consider the
explicit form of such counter terms.

\subsection{\label{PauliCT} Pauli-like counter terms}
In this section, we consider the explicit form of counter terms corresponding to diagrams N3 and N5 in Fig.~\ref{Sample}.   Counter terms corresponding to diagram N4 are those already appearing in the EFT for pions only.   At NNLO, examples of the LECs associated with N4-type counter terms are
the $\co(p^4)$ LEC $\ell_6$ (for its role in the ChPT approach to $a_\m^{\rm HVP}$ see Ref.~\cite{us2}) and the $\co(p^6)$ LEC $c_{56}$ \cite{BCE} (see App.~\ref{timemom}); their role is to renormalize the HVP subdiagram of
diagram N1$^{\rm EFT}$ in Fig.~\ref{Sample}.   Here, instead, we will construct 
the simplest example of a counter tem in the EFT which includes also muons and photons.

At N$^3$LO, $\hat\P(q^2)\sim q^4$ modulo logarithmic corrections, and, since $f(q^2)\sim m_\m^4/q^6$, the integral in Eq.~(\ref{muonan}) diverges
at this order, requiring an N5-type counter term.   Of course, the counter term
that is needed must be proportional to
\begin{equation}
\label{Pauli}
\psibar_\m\, \s_{\k\l}\, \psi_\m F_{\k\l}\ ,
\end{equation}
where $\psi_\m$ is the muon field, and $F_{\k\l}$ is the electromagnetic field
strength.   However, we will also need N3-type counter terms, and in general,
in the theory with pions, muons and photons, we need to analyze the general structure N3- and N5-type counter terms can have.   The standard method for carrying out this analysis is through the use of spurions.

We start with coupling massless, two-flavor QCD to $[SU(2)\times U(1)]_L
\times[SU(2)\times U(1)]_R$ vector sources $\ell_\k$ and $r_\k$, which
we will eventually set equal to the photon field $A_\k$ by choosing
\begin{equation}
\label{spurions}
\ell_\k=-\half(1-\t_3)A_\k\ ,\qquad r_\k=-\half(1-\t_3)A_\k\ ,
\end{equation}
where $\t_i$ are the Pauli matrices.   We also introduce a muon doublet,
$\psi=(\psi_\n,\psi_\m)^T$, which couples to the spurions $\ell_\k$ and $r_\k$
through
\begin{equation}
\label{muon}
\bar\psi\g_\k(\partial_\k+i\ell_\k P_L+i r_\k P_R)\psi\ ,
\end{equation}
where $P_R$ and $P_L$ are right- and left-handed projectors.   The
non-linear pion field $U=\mbox{exp}[2i\p/f_\p]$ couples to the spurions
through the covariant derivative
\begin{equation}
\label{covpi}
D_\k U=\partial_\k U+i\ell_\k U-iUr_\k\ .
\end{equation}
Since, for large momenta, the weight $f(q^2)$ is proportional to $m_\m^4$,
the N3- and N5-type counter terms have to contain the third power of
the muon mass,\footnote{One factor $m_\m$ in $f(q^2)$ comes from the
definition of $a_\m^{\rm HVP}$, and not from diagram N0.} and we thus need
to introduce a spurion $\c^{(\m)}$ for the muon mass as well.   Finally,
since $a_\m^{\rm HVP}$ contains internal photon lines, we need the charge 
matrix spurions $Q_L$ and $Q_R$ through which the photon couples to the
left-handed and right-handed quarks, respectively.   Of course, 
$Q_L=Q_R=Q=\mbox{diag}(\frac{2}{3},-\frac{1}{3})$, but the spurions
$Q_L$ and $Q_R$ transform differently, under $[SU(2)\times U(1)]_L$ and
$[SU(2)\times U(1)]_R]$, respectively.
QCD
coupled to the muon doublet $\psi$ and the spurions $\ell_\k$, $r_\k$
and $\c^{(\m)}$, and thus our EFT, is invariant under\footnote{We ignore the $U(1)$ axial
anomaly, because the corresponding source will be set equal to zero.}
\begin{eqnarray}
\label{transf}
U&\to& LUR^\dagger\ ,\\
\psi&\to& (LP_L+RP_R)\psi\ ,\nonumber\\
\ell_\k&\to& L\ell_k L^\dagger-iL\partial_\k L^\dagger\ ,\nonumber\\
r_\k&\to& Rr_k R^\dagger-iR\partial_\k R^\dagger\ ,\nonumber\\
\c^{(\m)}&\to& L\c^{(\m)}R^\dagger\ ,\nonumber\\
Q_L&\to& LQ_LL^\dagger\ ,\nonumber\\
Q_R&\to& RQ_RR^\dagger\ ,\nonumber
\end{eqnarray}
where $L\in [SU(2)\times U(1)]_L$ and $R\in [SU(2)\times U(1)]_R$.
For a complete construction of the EFT also a spurion $\c^{(\p)}$ for the
quark mass transforming in the same way as $\c^{(\m)}$ would be needed, but we will not need it for the counter
terms discussed below.

Ignoring the pions for the moment, the simplest counter term leading to
the Pauli structure~(\ref{Pauli}) is
\begin{equation}
\label{Paulimmass}
\bar\psi_L\s_{\k\l}\c^{(\m)} r_{\k\l}\psi_R+\bar\psi_R\s_{\k\l}r_{\k\l}\c^{(\m)\dagger}\psi_L
+\bar\psi_R \s_{\k\l}\c^{(\m)\dagger}\ell_{\k\l}\psi_L+\bar\psi_L\s_{\k\l} \ell_{\k\l}\c^{(\m)}\psi_R\ .
\end{equation}
At least one power of $\c^{(\m)}$ is needed, consistent with the fact that
one factor $m_\m$ has to appear because of the helicity flip of the muon
associated with its magnetic moment.   Since we need the third power
of the muon mass, two more spurion factors $\c^{(\m)}$ or $\c^{(\m)\dagger}$
need to be inserted.   This can be done in various ways consistent with the
symmetry~(\ref{transf}), but they all collapse to the same factor $m_\m^3$
once we set $\c^{(\m)}=m_\m$.

The invariant operator with the lowest dimension involving the pion field $U$ is
\begin{equation}
\label{pionfactor}
\tr(Q_LUQ_RU^\dagger)\ ,
\end{equation}
where $Q_L$ and $Q_R$ appear because of the internal photon lines in diagram N0.   Multiplying Eq.~(\ref{Paulimmass}) with two more insertions of
the spurion $\c^{(\m)}$ with Eq.~(\ref{pionfactor}), setting $\c^{(\m)}=m_\m$,
$\ell_\k$ and $r_\k$ equal to the values in Eq.~(\ref{spurions}) and
$Q_L=Q_R=Q$, we obtain the
counter term
\begin{equation}
\label{N3N5}
\frac{\a^2m^3_\m}{(4\p f_\p)^4}\, F_{\k\l}\psibar_\m\s_{\k\l}\psi_\m\,\tr\!\!\left(Q_LUQ_RU^\dagger\right)
=\frac{\a^2 m^3_\m}{(4\p f_\p)^4}\, F_{\k\l}\psibar_\m\s_{\k\l}\psi_\m\!\left(\frac{5}{9}-\frac{4}{f_\p^2}\,\p^+\p^-+
\co(\p^4)\!\right).
\end{equation}
We multiplied with the factor $1/(4\p f_\p)^4$ to make this a dimension-four
operator, with the scale $4\p f_\p$ standing in for the
hadronic scale $M_H$ of the ``heavy'' hadrons that have been integrated
out. The two powers of $\a$ reflect the presence of the two internal photon lines. The fact that the pion fields are traced over corresponds to the fact that they appear
in a loop; if there were $n$ pairs of charged pions, their contribution to $a_\m^{\rm HVP}$ would be
$n$ times larger.\footnote{Indeed, other charged mesons, such as the $K^+K^-$ pair, do
contribute, but their contribution is suppressed because of the larger mass of these mesons.} Furthermore, it is clear that the square of the quark charge matrix has to appear from the quark picture of the HVP.

Setting the pion field $\p=0$ in Eq.~(\ref{N3N5}), this counter term is of the
form~(\ref{Pauli}), and thus is the simplest example of an N5-type
counter term, with its coefficient renormalizing the overall divergence
that can appear in the EFT calculation of Eq.~(\ref{muonan}).   Since
such a divergence appears for the first time at N$^3$LO, we
expect this counter term to be of order $1/(4\p f_\p)^4$.   Indeed, dimensional
analysis leads to the appearance of this factor in Eq.~(\ref{N3N5}), and the
counter term thus takes a natural form.   Of course, divergences also
appear beyond N$^3$LO, and corresponding counter terms
proportional to powers of $1/(4\p f_\p)$ larger than four will be needed
as well.   Such counter terms are easily constructed by inserting
(covariant) derivatives and/or powers of the pion mass.

The counter term~(\ref{N3N5}) also produces a counter term of order $1/(4\p f_\p)^6$ with a photon,
two muon and two pion external lines, and such a counter term leads to diagrams of type
N3 in Fig.~\ref{Sample}.  The power of $1/(4\p f_\p)$ thus suggests that
this counter term will only be needed at N$^4$LO.   Moreover,
only N3-type (and not N5-type) counter terms will contribute to
pion-induced FV effects.   This is consistent if the first UV-divergent
FV effects requiring an N3-type counter term only appear at N$^4$LO, so that they are renormalized
by the N3-type counter term in Eq.~(\ref{N3N5}).    As we will see in Sec.~\ref{toymodel},
in order to produce a FV effect, a pion line needs to be cut.
This makes it plausible that no UV-divergent FV effects occur at three-loop
order, because cutting one pion line reduces the degree of divergence.
While a full three-loop calculation is beyond the scope of this paper,
in the next subsection we give a more quantitative argument supporting
this conjecture.

\subsection{\label{example} An explicit example}
A very simplified model for the ``three-loop" HVP is given by
\begin{equation}
\label{onet}
\P_0(q^2)=\frac{q^4}{f_\p^4}\int d^4p\,\frac{1}{p^2+m_\p^2}\frac{1}{(p-q)^2+m_\p^2}\ .
\end{equation}
Obviously this is clearly not a true three-loop contribution. However, in accordance with Eq.~(\ref{ChPTdiv}),  this model ``three-loop'' HVP  behaves like $\mathcal{O}(q^4 \log q^2)$ and is divergent, and these are the essential ingredients we need for our discussion.\footnote{Of course, at three loops more
complicated logarithmic corrections to the $q^4$ behavior can appear,
but we believe this is not essential to the point we wish to make.}

Regulating the integral as  $d^4p\to dp^2 p^2(p^2/\m^2)^\e$, thus introducing the unphysical scale $\m$, and ignoring numerical factors from angle
integrations, \etc, the result is
\begin{eqnarray}
\P_0(q^2)&=&
\frac{q^4}{f_\p^4}\int_0^1 dx \left(-1-\frac{1}{\e}-\log\frac{x(1-x)q^2+m_\p^2}{\m^2}\right)\ \nonumber \\
&=&\frac{q^4}{f_\p^4}\left(-\frac{1}{\e}+1-\log\frac{m_\p^2}{\m^2}
+\frac{\sqrt{4m_\p^2+q^2}}{q}\,\log\frac{\sqrt{4m_\p^2+q^2}-q}{\sqrt{4m_\p^2+q^2}+q}\right) . \label{twot}
\end{eqnarray}
Adding the ``ChPT'' counter term
\begin{equation}
\label{3threet}
\Pi_{\rm CT}(q^2)=\frac{q^4}{f_\p^4}\left( \frac{1}{\e}+ \ell(\m)-1 \right)\ ,
\end{equation}
in order to subtract the divergence in Eq.~(\ref{twot}), and allowing for an additional 
finite renormalization $\frac{q^4}{f_\p^4}(\ell(\m)-1)$,
one obtains the renormalized ``three-loop" pion vacuum polarization
\begin{eqnarray}
\label{fourt}
\hat{\P}_0(q^2)&\equiv&\P_0(q^2)+\Pi_{\rm CT}(q^2)\\
&=&\frac{q^4}{f_\p^4}\left(\ell(\m) - \int_0^1 dx \ \log \frac{x(1-x)q^2+m_\p^2}{\m^2}\right) \nonumber\\
&=&\frac{q^4}{f_\p^4}\left(\ell(\m)-\log\frac{m_\p^2}{\m^2}+  \frac{\sqrt{4m_\p^2+q^2}}{q}\,\log\frac{\sqrt{4m_\p^2+q^2}-q}{\sqrt{4m_\p^2+q^2}+q} \right)\ .\nonumber
\end{eqnarray}
The renormalized quantity $\hat{\P}_0(q^2)$ is a physical quantity, and thus should
not depend on the unphysical scale $\m$.
Therefore, in this simple example the running of the renormalized LEC $\ell(\mu)$ satisfies
\begin{equation}
\label{fourat}
\m\frac{d}{d\m}\,\ell(\m)=-2\ .
\end{equation}
Since the model of Eq.~(\ref{onet}) is not UV complete, we cannot determine the dependence of $\ell(\m)$ on the UV physics that has been integrated out, and the running of $\ell(\m)$ in (\ref{fourat}) is all we can know. In Sec.~\ref{toymodel} we will study a simple model which is UV complete.

As a model for the contribution to the muon anomaly from $\hat{\P}_0(q^2)$ we will choose the model ``anomaly'' $a$ to be given by the integral ($m\equiv m_\m$)
\begin{equation}
\label{fivet}
a=m^4 \int\frac{d^nq}{q^2+m^2}\ \frac{1}{q^6}\ \hat{\P}_0(q^2) \left(\frac{M^2}{q^2+M^2}\right)\ ,
\end{equation}
where $1/q^6$ represents the combination of  photon propagators and other kinematical factors and $1/(q^2+m^2)$ regulates the IR divergence (\ie, it gives the dependence on the muon mass in our simplified representation).   The factor $M^2/(q^2+M^2)$ has been inserted to make the
$q$ integral in Eq.~(\ref{fivet}) finite, and ``stands in'' for the non-pionic hadron physics of QCD.   One might think of it as the insertion of a fake $\r$ propagator, but this is not essential: the only job of this factor is to regulate the UV divergence of the $q$ integral.   As we will see, it will allow us to determine the form of the counter term needed to
make $a$ finite without this factor.

As it stands, the integral~(\ref{fivet}) is finite and well behaved for $d=4$. The   ``ChPT'' version of $a$ is obtained by setting  $M^2/(q^2+M^2)\to 1$, \ie, sending $M^2\to \infty$, but of
course, that reintroduces the UV divergence of the $q$ integral.

In order to proceed, we split
\begin{equation}
\label{sixt}
\frac{M^2}{q^2+M^2}=1-\frac{q^2}{q^2+M^2}\ ,
\end{equation}
and using the expression for $\hat{\P}_0(q^2)$ in Eq.~(\ref{fourt}) one may split the contributions to $a$ as
\begin{subequations}
\label{sevent}
\begin{eqnarray}
a&=&a_{\rm EFT}+a_{\rm CT} \nonumber \\
&=&\frac{m^4}{f_\p^4}\int_0^\infty \frac{dq^2}{q^2+m^2}\left(\frac{q^2}{\m^2}\right)^\e\left( \ell(\m)-\int_0^1 dx \log \frac{q^2 x (1-x) + m_\p^2}{\mu^2}  \right) \label{seventa}\\
&& \hspace{-2.5cm}  - \ \frac{m^4}{f_\p^4}\int_0^\infty \frac{dq^2}{q^2+\underbrace{m^2}_{neglect}}\ \!\!\left(\frac{q^2}{\m^2}\right)^\e\!\!\frac{q^2}{q^2+M^2}\left( \ell(\m)-\int_0^1 dx \log \frac{q^2 x (1-x) + m_\p^2}{\mu^2} \right)\label{seventb},
\end{eqnarray}
\end{subequations}
 where, as indicated, the muon mass $m$ may be neglected in the second integral, $a_{\rm CT}$, on account of the extra $q^2$ in the numerator. As we will see, this also allows the limit $m_\p^2 \to 0$ to be taken, rendering $a_{\rm CT}$ independent of the IR scales $m$ and $m_\p$, as expected
for a counter term.

The integrals yield cumbersome expressions for $m\ne m_\p$. This is why we will simplify our example by taking  $m=m_\p$  as a common low-energy scale, $m=m_\p \ll M$, in the rest of this section.   This simplification only serves to simplify the math and is not essential, of course.

Setting $m_\p=m$,  the result for $a_{\rm EFT}$ can be obtained by
evaluating the integrals in Eq.~(\ref{seventa}),
 \begin{eqnarray}
 a_{\rm EFT}&=&\frac{m^4}{f_\p^4}\left[ \ell(\m) \left(-\frac{1}{\e}- \log\frac{m^2}{\m^2} \right) \right. \nonumber \\
 &&  \left. - \frac{1}{\e^2}- \frac{2}{\e}- 2 \log \frac{m^2}{\m^2}
+ \frac{1}{2}\log^2\frac{m^2}{\m^2} + \frac{2}{27}\,\p^2+ \cc \right]\ , \label{ninet}
 \end{eqnarray}
 where the first (second) line corresponds to the first (second) term in the integrand of Eq.~(\ref{seventa}), and $\mathcal{C}$ is a constant given by
 \begin{equation}
 \mathcal{C}=-\frac{1}{9}\,\psi'\left(\frac{1}{6}\right)-\frac{1}{18}\,\psi'\left(\frac{1}{3}\right)
 +\frac{1}{9}\,\psi'\left(\frac{2}{3}\right)
 +\frac{1}{18}\,\psi'\left(\frac{5}{6}\right) \ ,\label{ninetb}
 \end{equation}
 where $\psi'(x)=\frac{d\psi(x)}{dx}$ with $\psi(x)$ is the digamma function $\psi(x)=\frac{\G'(x)}{\G(x)}$.

 We next turn to $a_{\rm CT}$ in Eq.~(\ref{seventb}). The result of the integral is
 \begin{eqnarray}
 a_{\rm CT}&=&-\ \frac{m^4}{f_\p^4}\left[ \ell(\m)\left( - \frac{1}{\e}- \log\frac{M^2}{\m^2}\right) \right.\nonumber \\
&&- \frac{1}{\e^2}-\frac{2}{\e}- 2\log\frac{M^2}{\m^2}+ \frac{1}{2}\log^2\frac{M^2}{\m^2}+\frac{\p^2}{6}\nonumber \\
&&\hspace{3cm} + \left. \mathcal{O}\left(\frac{m_\p^2}{M^2}\log^2\frac{m_\p^2}{M^2}  \right)\right]
 \ , \label{tent}
\end{eqnarray}
where again the first (second)  line corresponds to the first (second) term in the integrand of Eq.~(\ref{seventb}). We emphasize that for the second line,
\ie, the term  with the $\log\left(q^2 x (1-x) + m_\p^2\right)$, the limit $m_\p^2\to 0$ may be taken, resulting in a term independent of $m_\pi^2$.\footnote{Dependence on $m_\p$ would be provided by higher-order counter terms.} In the third line of Eq.~(\ref{tent}) we have kept $m_\p^2$ explicit for illustration, but this whole term is $ \mathcal{O}\left(\frac{m_\p^2}{M^2} \right)$ and therefore is to be neglected at the order we consider. In App.~\ref{integrals} we show how to calculate it from Eq.~(\ref{seventb}).

Adding  $a_{\rm EFT}$ and $a_{\rm CT}$ in Eqs.~(\ref{ninet}) and~(\ref{tent}) we finally obtain the result
\begin{equation}
a=\frac{m^4}{f_\p^4} \left( \ell(\m) \log \frac{M^2}{m^2}+ 2 \log \frac{M^2}{m^2}+ \frac{1}{2}\log\frac{M^2}{m^2}\log\frac{\m^4}{m^2M^2}-\frac{5}{54}\p^2+ \mathcal{C}\right)\  ,\label{elevent}
\end{equation}
where $\mathcal{C}$ is given in Eq.~(\ref{ninetb}). Note how the condition~(\ref{fourat}) for the running of $\ell(\m)$ makes our result
independent of $\m$.

The result for $a_{\rm CT}$ in Eq.~(\ref{tent}) admits an expansion in powers of $m_\p^2/M^2$. To lowest order in this expansion, there are no logarithms such as $\log m_\p^2/M^2$, which could only come from a pion loop. Therefore, the conclusion is that the leading order is given by an N5-type Pauli-like operator, without any pion loops.   Thus at leading order
(which, recall, in the real world corresponds to three-loop order), only an N5-type counter term is needed.
It is only at the next order in the chiral expansion, \ie, $\co(m_\p^2/M^2\times m_\m^4/f_\p^4)$, that these logarithms appear: this is when tadpole diagrams of type N3 will start to contribute.

\subsection{\label{exfinvol} The example in finite volume}
Let us now extend the discussion of our example to have a first look at FV effects coming
from the pions.   We thus want to consider the integral
\begin{equation}
\label{3one}
D(q^2)=\int_{-\infty}^{\infty}\frac{dp_4}{2\p}\int\frac{d^3p}{(2\p)^3}\
\frac{1}{p_4^2+p^2+m_\p^2}\ \frac{1}{(p_4-q_4)^2+p^2+m_\p^2}
\end{equation}
in a finite spatial volume $V=L^3$ of linear dimension $L$, with periodic boundary
conditions.   Here we took $q=(0,0,0,q_4)$ to point in the 4-direction, without loss
of generality, as this will be convenient in our explicit calculations.   In finite volume,
the integral over $\vec p$ is replaced by a sum, and in finite volume $D$ becomes
\begin{equation}
\label{3oneFV}
D_{\rm FV}(q_4^2)=\int_{-\infty}^{\infty}\frac{dp_4}{2\p}\frac{1}{L^3}\sum_{\vec p=2\p\vec n/L}\
\frac{1}{p_4^2+p^2+m_\p^2}\ \frac{1}{(p_4-q_4)^2+p^2+m_\p^2}\ ,
\end{equation}
where $\vec n$ has integer components.  In order to isolate the FV effects, we will use Poisson resummation, 
\begin{equation}
\label{Afourteen}
\frac{1}{L^3}\sum_{\vec{p}} f(\vec{p}^2)=\sum_{\vec{n}}\int \frac{d^3\vec{p}}{(2\p)^3}\ e^{i\vec{n}\cdot\vec{p}\, L} f(\vec{p}^2) = \sum_{\vec{n}}\left( \frac{1}{4i\p^2n L} \right)\int_{-\infty}^\infty dp\, p\,f(p^2) e^{inpL}\ ,
\end{equation}
in which $n=|\vec n|$.   In position space, the vector
$\vec n$ represents the number of times the pion wraps around the periodic spatial volume 
in each direction.  Thus, the term with $n=0$ corresponds to the infinite-volume
part, and the $n>0$ terms separate out the FV contributions.  
Using Eq.~(\ref{Afourteen}),
we find that
\begin{eqnarray}
\label{DFV}
D_{\rm FV}(q_4^2)&=&\sum_{\vec n} D^{(n)}(q_4^2)\ ,\\
D^{(n)}(q_4^2)&=& \int_{-\infty}^{\infty}\frac{dp_4}{2\p}\int_{-\infty}^{\infty}\frac{dp}{2\p^2}\, p\,  \frac{1}{p_4^2+p^2+m_\p^2}\ \frac{1}{(p_4-q_4)^2+p^2+m_\p^2}\,\frac{e^{i np L}}{2 i nL}\nonumber\\
&=&\frac{1}{8 \p^2 q_4 nL}  \int_{0}^{\infty}\frac{dp_4}{p_4} \left(e^{-nL\sqrt{(p_4-q_4/2)^2+m_\p^2}}- e^{-nL\sqrt{(p_4+q_4/2)^2+m_\p^2}}\right)\ ,\nonumber
\end{eqnarray}
where we used contour integration to evaluate the integral over $p$, and we shifted $p_4\to p_4+q_4/2$.   For large $m_\p L$, the integral over $p_4$ is (exponentially) dominated by the region
$(p_4\pm q_4/2)^2\ll m_\p^2$, allowing us to expand the square roots for large $m_\p^2$.
For large $m_\p L$ and $n>0$ we thus approximate
\begin{eqnarray}
D^{(n)}(q_4^2)&\approx &\frac{1}{8 \p^2 q_4 nL}\ e^{-m_\p nL} \int_{0}^{\infty}\frac{dp_4}{p_4}\left( e^{-\frac{nL m_\p}{2}\, \frac{(p_4-q_4/2)^2}{m_\p^2}}- (q_4\to -q_4) \right)\nonumber \\
&= & \frac{1}{8 \p^2 q_4 nL}\ e^{-m_\p nL} \,\p\, e^{-\frac{q_4^2 nL}{8 m_\p}}\ \mathrm{Erfi}\left(  \sqrt{\frac{q_4^2 nL}{8 m_\p}}\right)\ ,\label{3three}
\end{eqnarray}
where $\mathrm{Erfi}(x)$ is the imaginary error function $\mbox{erf}(ix)/i$, which is real for real $x$.\footnote{In order to carry out the integral in Eq.~(\ref{3three}), we regulated the pole at $p_4=0$ by introducing a factor $p_4^\e$ in each term, then combined both terms and only at the end took $\e\to 0$.}

Using the asymptotic expansion
\begin{equation}
\label{Erfiexp}
\mathrm{Erfi}(x)\approx e^{x^2}\left( \frac{1}{\sqrt{\p} x }+ \frac{1}{2\sqrt{\p} x^3} + \frac{3}{4\sqrt{\p}x^5}+ ... \right) \ ,\qquad  (x\to \infty)\ ,
\end{equation}
we obtain, for $n>0$,
\begin{equation}
\label{Das}
D^{(n)}(q_4^2)\approx \frac{\sqrt{2\p}}{4\p^2}\ \frac{e^{-m_\p nL}}{(q_4 nL)^2}\sqrt{m_\p nL}\left( 1+ \mathcal{O}\left(\frac{1}{m_\p nL};\frac{m_\p nL}{(q_4 nL)^2}\right)\right)\ .
\end{equation}
We recall that $\P_0(q^2)= \frac{q^4}{f_\p^4}\,D(q^2)$, {\it cf.} Eq.~(\ref{onet}). Consequently, we see here an example of what we anticipated earlier in this section about the N$^3$LO contribution to $a_\m^{\rm HVP}$. The insertion of $\hat{\P}_0(q^2)$ into the $a$ integral, as in Eq.~(\ref{fivet}), leads to a divergence in infinite volume which is renormalized by the corresponding counter term, {\it cf.} Eqs.~(\ref{seventa}),~(\ref{seventb}). On the other hand, the finite-volume contribution, Eq.~(\ref{Das}), when inserted into the integral~(\ref{fivet}) for $a$ is UV finite even for $M\to \infty$.\footnote{Equation~(\ref{Das}) is only valid for large $q_4L$.   This is sufficient for our argument, as it is the large-$q$ region of the integral over $q$ in Eq.~(\ref{fivet}) that leadsto a UV divergence.}
 
Intuitively, this can be understood as follows.   To obtain the leading FV correction, we take the
one pion loop in any diagram to wrap ``around the world.''   Such a pion can be seen as an on-shell pion (as we will show explicitly in Sec.~\ref{toymodel}).   Effectively, one thus
``removes" a loop and a pion propagator, decreasing the degree of divergence by (at least) two.

However, as we have discussed above, we expect that FV effects in the real world will lead to a divergent $a_\m^{\rm HVP}$\ integral at N$^4$LO. Therefore, in the next section, we will study an even simpler example which, while keeping integrals at an elementary level, does lead to divergent FV effects and allows us to illustrate the interplay of these FV effects with the corresponding UV counter terms which appear at infinite volume.

\section{\label{toymodel} A toy model}
Our goal is to investigate the interplay between UV divergences, renormalization, and FV effects in more detail.   As we have argued, this interplay shows up only in full force
at three and four loops in the case of $a_\m^{\rm HVP}$.    Therefore, in this section, we will study this
interplay in a very simple model, in which we do not have to go beyond two loops in order to
see this interplay at work, and in which details have been kept simple enough to make
explicit calculations feasible.

We will define the model in Sec.~\ref{model}, where we explain how keeping things simple
led us to consider this model.   Then, in Sec.~\ref{infvol}, we will essentially repeat the
analysis of Sec.~\ref{CTFV} for the model, in infinite volume.   In Sec.~\ref{finvol} we will
demonstrate how, once the infinite-volume counter terms have been identified,
FV corrections due to the ``pions'' in our model are always UV finite, and thus
well defined.

\subsection{\label{model} Definition of the model}
\begin{figure}[t]
\begin{center}
\includegraphics[width=3in]{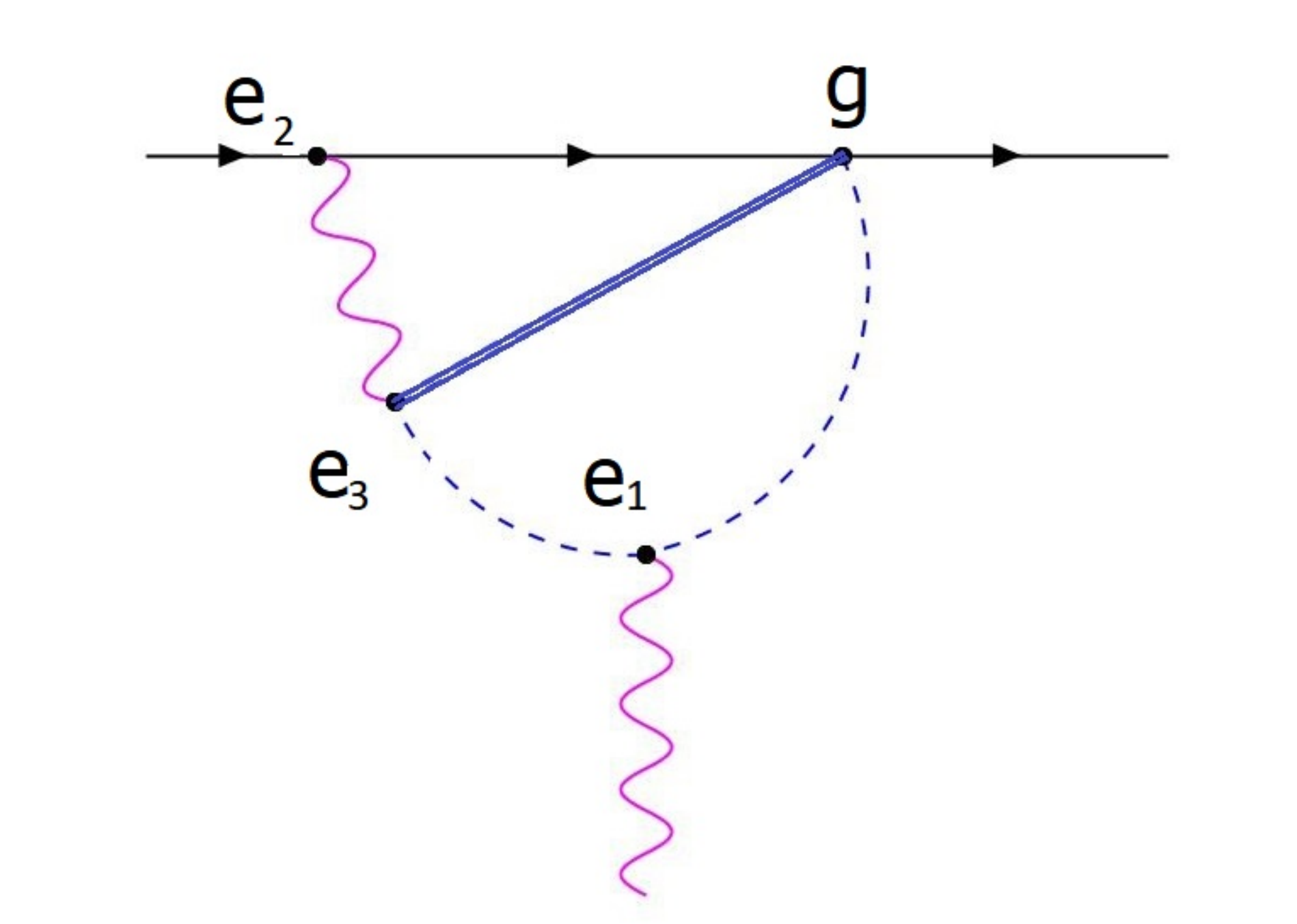}
\floatcaption{fig:figa}{Feynman diagram for the quantity $a$ defined in Eq.~(\ref{one}).}
\end{center}
\end{figure}
We wish to construct a simple toy model with pions, muons, photons, and some
``heavy strong-interaction'' physics in which a quantity $a$ analogous to $a_\m$ can
be defined.
While we really just need an integral like the one in Eq.~(\ref{muonan}),
it is instructive to cast the model in terms of a lagrangian and Feynman rules
obtained from the lagrangian.   This will allow for a diagrammatic analysis analogous
to that based on Fig.~\ref{Sample}.

The lagrangian for our model is\footnote{All similarity with the linear sigma model is purely coincidental.}
\begin{eqnarray}
\label{modelL}
\cl&=&\half(\partial_\k\p)^2+\half m_\p^2\p^2+\half(\partial_\k\s)^2+\half m_\s^2\s^2\\
&&+\half(\partial_\k\psi_\m)^2+\half m_\m^2\psi_\m^2+\half(\partial_\k A)^2\nonumber\\
&&+\half\, e_1 A\p^2+\half\, e_2 A\psi_\m^2+e_3 A\s\p+\half\,g\psi_\m^2\s\p\ .\nonumber
\end{eqnarray}
All fields are scalars, but we can intuitively think of the massless scalar $A$ as a ``photon,''
the scalar $\psi_\m$ as a ``muon,'' while of course $\p$ is our ``pion.''\footnote{All couplings in Eq.~(\ref{modelL}) except $g$ have mass dimension one.} The strong-interaction
physics is represented by the massive scalar $\s$, and we will thus always think of $m_\s$
as much larger than $m_\p$ and $m_\m$.

\begin{figure}[!ht]
\begin{center}
\includegraphics[width=5in]{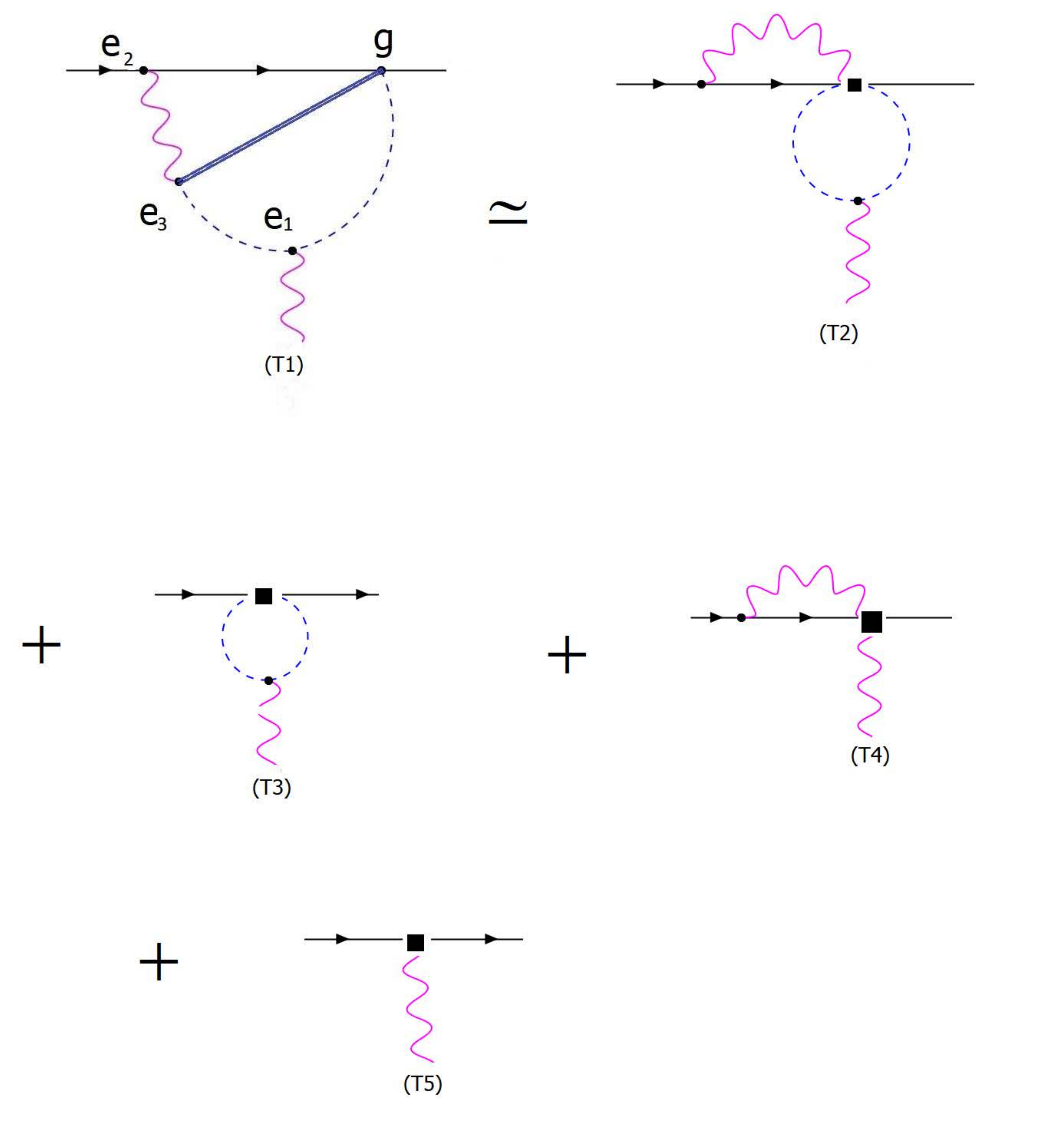}
\floatcaption{Sample2}{$a_\m^{\rm HVP}$ for the toy model.}
\end{center}
\end{figure}

In this model, we define a ``muon anomaly'' by
\begin{equation}
\label{one}
a=e_1e_2e_3g\int_0^\infty dq^2\, \frac{m_\m^2}{m_\m^2+q^2}\, \Pi(q^2)\ ,
\end{equation}
where, as in Eq.~(\ref{muonan}), $q^2>0$ is euclidean.  Here we omitted purely numerical
factors from the integration over the angles of $q$, and a possible symmetry factor.
In Secs.~\ref{infvol} and~\ref{finvol} we will omit the factor $e_1e_2e_3g$ as well.
The function $\Pi(q^2)$ is given by 
\begin{equation}
\label{two}
\Pi(q^2)=\int\frac{d^4 p}{(2\pi)^4}\, \frac{1}{(p^2+m_\s^2)}\frac{1}{\left[ (p-q)^2+m_\p^2 \right]^2}\ ,
\end{equation}
which can be seen by traversing around the hadronic loop in the diagram depicted in Fig.~\ref{fig:figa}.
We emphasize that this model for $\P(q^2)$ has been chosen to be UV finite without
any further subtraction, unlike Eq.~(\ref{subvacpol}).   In particular, this means that also $\P(0)$
is finite in the model.   We could have chosen a model with $\P(q^2)$ defined by Eq.~(\ref{two})
without the square on the pion propagator, in which case $\P(q^2)$ would have been
logarithmically divergent.   In that case, we would have considered the UV-finite
difference $\P(q^2)-\P(0)$, as in Eq.~(\ref{muonan}).   However, this would make the 
mathematical treatment of the model more cumbersome, and it is not essential, as we will
see next.   It is thus important to keep in mind, for the rest of this section, that $\P(q^2)$
itself is finite, and does not need to be subtracted, unlike in the real world.

The diagram for the ``anomaly'' $a$ is shown in Fig.~\ref{fig:figa}.  In detail, it looks
rather different from diagram N0 in Fig.~\ref{Sample}.\footnote{In fact, it bears some
similarity to the hadronic light-by-light contribution to $a_\m$, with the external
photon line attached to the pion.}  But, it shares the following essential properties with $a_\m^{\rm HVP}$.   First, $a$ is a UV-finite quantity, just like $a_\m^{\rm HVP}$,
and it is an integral over a weight function times a hadronic loop, $\P(q^2)$, which itself
is finite, just like $\hat\P(q^2)$ in Eq.~(\ref{muonan}).   However, as can be seen from
Fig.~\ref{fig:figa}, if we ``integrate out'' the $\s$, \ie, we contract its propagator to a
point by replacing $1/(p^2+m_\s^2)\to 1/m_\s^2$ in Eq.~(\ref{two}), both the pion loop
and the muon-photon loop become logarithmically divergent.  This can be seen in
diagram T2 of Fig.~\ref{Sample2}.
This implies that in
an EFT containing only the pions, photons and muons, while the $\s$ has been integrated
out, counter terms will need
to be introduced to renormalize these divergences.   Counter terms will be needed
for the pion-loop subdivergence (diagram T4), for the muon-photon-loop subdivergence
(diagram T3), and the overall two-loop divergence (diagram T5).   The precise form
of these counter terms will be derived in Sec.~\ref{infvol} below.

The example of the quantity $a$ in the toy model thus mimics the situation that arises at N$^3$LO
in the case of $a_\m^{\rm HVP}$.  
At N$^3$LO in ChPT not only are counter terms needed
to renormalize UV divergences in the HVP, but also the integral over $q^2$ (the momentum through the muon and photon lines in
Eq.~(\ref{muonan})) which
becomes divergent and leads to the counter terms discussed in Sec.~\ref{CTFV}.
Because of the construction of our model, and the definition of the quantity $a$,
the same phenomena happen here already at the lowest possible number of
loops.  Since we want to study in our model both ``strong
interaction'' counter terms of type N4 and ``electromagnetic'' counter terms
of type N3 and N5, we need a subdivergence from a pion loop, and a subdivergence
from a loop containing a muon, and both are present in diagram T2 of Fig.~\ref{Sample2}, which represents the EFT version of the diagram in Fig.~\ref{fig:figa} obtained by
contracting the $\s$ propagator to a point.   
The detailed form of $\P(q^2)$ makes the example
somewhat contrived, and leads to the detailed form of the diagram in Fig.~\ref{fig:figa}
and thus the diagrams in Fig.~\ref{Sample2} to be different from the diagrams in
Fig.~\ref{Sample}.   We emphasize that this is not important; what is important is
the fact that the model generates both types of counter terms already at two loops.
In the next subsection, we will carry out the integrals, transition to the EFT, and
fully elucidate the counter-term structure in the toy model.

We end this subsection with a comment.   In the model~(\ref{modelL}), there are
other contributions to the quantity $a$, \ie, other diagrams with 
an external photon and two external muon lines.   In particular, a contribution exists with one $\m^2\s\p$ and one
$A\s\p$ vertex, proportional to $e_3 g$, and this contribution is UV divergent.   But
this will not affect our discussion of the diagram in Fig.~\ref{fig:figa}.  While the EFT version of the simpler one-loop
contribution can be analyzed with the same method as we will use below to analyze the
integral in Eq.~(\ref{one}), it does not exhibit the parallel behavior with $a_\m^{\rm HVP}$\ we are after.
We will thus ignore this contribution, and define $a$ for the rest of this section
by the integral in Eq.~(\ref{one}).

\subsection{\label{infvol} The toy model in infinite volume}
We begin with the calculation of $a$ in the complete theory, Eq.~(\ref{modelL}).
Using Feynman parametrization of the integral in ~(\ref{two}), we find
(omitting, from now on, the couplings $e_{1,2,3}$ and $g$)
\begin{eqnarray}
\label{Piq}
\P(q^2)&=&\frac{1}{16\p^2}\int_0^1 dx\ \frac{x}{q^2x(1-x)+m_\s^2(1-x)+m_\p^2x}\\
&=&
 \frac{1}{16\p^2}\left(\frac{1}{m_\s^2+q^2}\,{\log\frac{m_\s^2}{m_\p^2}}+ \frac{m_\s^2-q^2}{q^2(m_\s^2+q^2)}\,\log\frac{m_\s^2}{m_\s^2+q^2}   \right)+ \mathcal{O}\left(\frac{m_\p^2}{m_\s^4}\right)
\ .\nonumber
\end{eqnarray}
We emphasize again that $\P(0)$ is finite, as can be seen by setting $q^2=0$ in 
Eq.~(\ref{Piq}).   No subtraction is needed. In fact, we note that, to 
leading order in an expansion in $1/m_\s^2$, $\P(q^2)=\P(0)$.
As indicated in Eq.~(\ref{Piq}), we will work to leading order in $m_\p^2/m_\s^2$, as this will
be sufficient for our purposes.  Inserting this into the integral defining $a$ in Eq.~(\ref{one}),
we find for $a$ the explicit result\footnote{One way to do the calculation is to first do
perform the integral over $q^2$ using the Feynman-integral representation for $\P(q^2)$
of Eq.~(\ref{Piq}) and then carry out the integral over $x$.}
\begin{equation}
\label{aexact}
a=\frac{m_\m^2}{16\p^2m_\s^2}\left[\left( 1- \log \frac{m_\s^2}{m_\p^2} \right) \log \frac{m_\m^2}{m_\s^2} -1 +\frac{\p^2}{3}+  \mathcal{O}\left( \frac{m_\m^2, m_\p^2}{m_\s^2} \right) \right]\  .
\end{equation}
This is the result that an EFT analysis of our model is expected to reproduce, and this
analysis is what we will turn to next.

The EFT for our model is obtained by integrating out the $\s$ field, which amounts to
replacing every $\s$ propagator in the theory by $1/m_\s^2$, thus making the exchange
of a $\s$ meson into a point vertex.   This corresponds to splitting the $\s$ propagator in
Eq.~(\ref{two}) as
\begin{equation}
\label{split}
\frac{1}{p^2+m_\s^2}=\frac{1}{m_\s^2}-\frac{p^2}{m_\s^2(p^2+m_\s^2)}\ .
\end{equation}
The EFT contribution corresponding to the first term on the right-hand side is depicted
in diagram T2 of Fig.~\ref{Sample2}.   To begin with, this replacement makes $\P(q^2)$
divergent, and a counter term renormalizing this divergence will thus have to be
introduced in the EFT.   This counter term should of course reproduce the contribution
from the second term on the right-hand side of Eq.~(\ref{split}), in an expansion in inverse
powers of $m_\s^2$.   The replacement also gives rise to the divergent muon-photon loop in diagram T2; we will return to this divergence below.

As a side comment, one could study also this model at higher orders in $m_\p^2/m_\s^2$
and $m_\m^2/m_\s^2$ by further expanding the second term on the right-hand side of
Eq.~(\ref{split}), using $-p^2/(m_\s^2(p^2+m_\s^2))=-p^2/m_\s^4+\co(p^4/m_\s^6)$.
But since our goal is to keep things mathematically as simple as possible, we will
only consider the expansion in inverse powers of $m_\s^2$ to leading order, in the rest
of this paper.

Replacing $d^4p\to \m^\e d^dp$ with $d=4-\e$ in Eq.~(\ref{two}) allows us to calculate the
EFT part of $\P(q^2)$, and we find
\begin{equation}
\label{PiEFT}
\P_{\rm EFT}(q^2)=\frac{1}{16\p^2m_\s^2}\left(\frac{2}{\e}-\g_E-\log\frac{m_\p^2}{4\p\m^2}\right)=\P_{\rm EFT}(0)\ .
\end{equation}
We note that this result reproduces the $-\log{m_\p^2}/(16\p^2 m_\s^2)$ of Eq.~(\ref{Piq}).
The toy model is the UV completion of our EFT, and thus the second term in Eq.~(\ref{split})
should lead us to the form of the counter term.   At $q=0$ we thus find the counter-term
contribution by replacing the $\s$ propagator in Eq.~(\ref{two}) by this second term:
\begin{equation}
\label{PiCT}
\P_{\rm CT}(0)=-\frac{1}{16\p^2m_\s^2}\!\!\left( \frac{2}{\epsilon}-\gamma_E-\log \frac{m_\s^2}{4\p\m^2}+1 +\co\left(\frac{m_\p^2}{m_\s^2}\right)\right)\ .
\end{equation}
The sum $\P_{\rm EFT}(0)+\P_{\rm CT}(0)$ reproduces $\P(0)$:
\begin{equation}
\label{PiqEFT}
\P(0)=\frac{1}{16\p^2m_\s^2}\left(\log\frac{m_\s^2}{m_\p^2}-1+\co\left(\frac{m_\p^2}{m_\s^2}\right)\right)\ .
\end{equation}

Of course, if we do not know the underlying
UV completion, we cannot calculate the contribution~(\ref{PiCT}), as the precise
form of the second term in Eq.~(\ref{split}) is not known if we only have the EFT.  But,
the divergent and logarithmic terms in $\Pi(0)_{\rm CT}$ can be inferred from those of $\Pi(0)_{\rm EFT}$ and dimensional analysis, as the sum of the two has to be finite and independent of $\m$.   This is precisely how a counter term is introduced in the EFT,
in order to absorb the divergence and the scale dependence that shows up in the calculation
of Eq.~(\ref{PiEFT}).   This does not determine the remaining finite part
(here the term $(-1+\g_E-\log{4\p})/(16\p^2m_\s^2)$) of the counter term.  If only the EFT
is known, the finite part gets replaced by an unknown finite constant of the
right dimension: the value of the renormalized LEC in a particular regularization scheme
(here, dimensional regularization with minimal subtraction).
The counter term discussed here takes the form
\begin{equation}
\label{T4CT}
T4:\qquad \frac{C_{\psi_\m^2A^2}(\m)}{4}\, \psi_\m^2A^2 \ ,
\end{equation}
and corresponds to the black square in diagram T4 in Fig. \ref{Sample2}.

We now return to the EFT calculation of $a$ in Eq.~(\ref{one}).   Diagram T2 has more divergences for which counter terms need to be introduced, and we aim to identify those
from splitting $a$ into a low-energy part $a_1$ and a counter-term part $a_2$, following the
same reasoning as above for $\P(q^2)$:   
\begin{eqnarray}
\label{twelve}
a&=&\int_0^\infty dq^2\ \left(\frac{q^2}{\m^2}\right)^\epsilon \frac{m_\m^2}{q^2+m_\m^2}\left\{ \P_{\rm EFT}(0)+\P_{\rm CT}(0)+\left[\P(q^2)-\P(0)\right]\right\}\\
&=& a_1+a_2\nonumber\\
&=&\underbrace{m_\m^2 \int_0^\infty \frac{dq^2}{q^2+m_\m^2}\ \left(\frac{q^2}{\m^2}\right)^\epsilon \left\{ \P_{\rm EFT}(0)+\P_{\rm CT}(0)\right\}}_{a_1}\nonumber\\
&&\hspace{2cm}+\underbrace{\  m_\m^2 \int_0^\infty \frac{dq^2}{q^2+\underbrace{m_\m^2}_{\mathrm{neglect}}}\ \left(\frac{q^2}{\m^2}\right)^\epsilon \left\{\P(q^2)-\P(0)\right\}}_{a_2}\ .\nonumber
\end{eqnarray}
Since we want to split $a$ into two divergent parts, $a_1$ and $a_2$, 
we introduced a regulating factor $(q^2/\m^2)^\e$, necessary to define each of the terms
$a_1$ and $a_2$ separately; of course, $a$ itself is finite in the limit $\e\to 0$.  We emphasize that this regulator is independent of the regulator used in the calculation of $\P_{\rm EFT}(0)$ and $\P_{\rm CT}(0)$ in Eqs.~(\ref{PiEFT}) and~(\ref{PiCT})
(the sum of which is finite); the use of dimensional regularization again here is just for calculational simplicity.   The $m_\m^2$ in the denominator of the integral defining $a_2$ may be neglected because $ \P(q^2)-\P(0) \sim q^2$ as $q^2\rightarrow 0$. Indeed, while $a_1$ will depend on $m_\m^2$ logarithmically, $a_2$ is analytic in $m_\m^2$ to ${\cal O}(m_\m^2)$, which is the order to which we have constructed the EFT.\footnote{It can be shown that counter terms are analytic in the low-energy scales order by order in the EFT expansion.}
The contribution $a_1$ is what one would obtain using the EFT calculation of $\P(q^2)$
(which in this example we carried out to lowest order, \ie, we obtained an EFT representation
of $\P(0)$, including the counter term contribution $\P_{\rm CT}(0)$).   However, the integral
over $q^2$ defining $a_1$ is itself divergent, and we thus will need to add new counter terms,
which are only defined in the theory containing not only the pion, but also the photon
and muon as dynamical fields, with the integral over $q^2$ representing the photon-muon loop.   Of course, since
in our model we have the exact expression~(\ref{one}), we know what $a_2$ is exactly,
but we will see how it corresponds to counter terms in the pion-muon-photon EFT below.

A direct calculation of the first term in Eq.~(\ref{twelve}) gives
\begin{equation}
\label{thirteen}
a_1= m_\m^2 \left\{ \P_{\rm EFT}(0)+\P_{\rm CT}(0)\right\}\left(- \frac{1}{\e}-\log \frac{m_\m^2}{\m^2} \right)\ ,
\end{equation}
showing explicitly that this contribution is divergent; this divergence comes from the muon-photon loop in diagram T2.   A direct calculation of the second term yields\footnote{One way
to do the calculation is to start from the first line of Eq.~(\ref{Piq}) for $\P(q^2)$, subtract $\P(0)$,
carry out the integral over $q^2$ in Eq.~(\ref{twelve}), and finally the integral over $x$ in
Eq.~(\ref{Piq}).}
\begin{equation}
\label{fourteen}
a_2= m_\m^2 \left\{ \Pi_{EFT}(0)+\Pi_{CT}(0)\right\}\left(\frac{1}{\epsilon}+\log \frac{m_\s^2}{\m^2} \right)+\left( \frac{m_\m^2}{16\p^2 m_\s^2} \right)\left(-1 +\frac{\p^2}{3}\right)\ ,
\end{equation}
so that the total result is
\begin{equation}
\label{fifteen}
a=a_1+a_2= m_\m^2 \Pi(0) \log \frac{m_\s^2}{m_\m^2}+\left(\frac{m_\m^2}{16\p^2m_\s^2}\right)\left(-1 +\frac{\p^2}{3}\right)\ .
\end{equation}
Using Eq.~(\ref{PiqEFT}) this equals the complete result,  Eq.~(\ref{aexact}).

Equation~(\ref{fourteen}) is the
result in the underlying theory, which we do not know if do not know the UV completion of the EFT.  However, the divergence in $a_1$ demands that we
add counter terms to the EFT that absorbs this divergence, yielding the contribution shown
as the first term in Eq.~(\ref{fourteen}).   Again,
the divergent and logarithmic terms in Eq.~(\ref{fourteen}) can be inferred from those in Eq.~(\ref{thirteen}) and dimensional analysis as the sum has to be finite and independent of $\m$.
The second term in Eq.~(\ref{fourteen}) again corresponds to a finite contribution,
which is unknown if we only have access to the EFT, and will thus be represented in the EFT
by renormalized LECs.

In order to disentangle the complete counter-term structure, we split the
different contributions as:
\begin{eqnarray}
\label{sixteen}
a&=&\underbrace{m_\m^2 \Pi_{\rm EFT}(0) \left(- \frac{1}{\epsilon}-\log \frac{m_\m^2}{\m^2} \right)}_{(1)} \\
&&\hspace{1cm}+ \underbrace{m_\m^2  \Pi_{\rm CT}(0) \left(- \frac{1}{\epsilon}-\log \frac{m_\m^2}{\m^2} \right)}_{(2)}\nonumber\\
&& \hspace{2cm}+\underbrace{m_\m^2   \Pi_{\rm EFT}(0) \left(\frac{1}{\epsilon}+\log \frac{m_\s^2}{\m^2} \right)}_{(3)}\nonumber \\
&& \hspace{3cm}+\underbrace{m_\m^2   \Pi_{\rm CT}(0) \left(\frac{1}{\epsilon}+\log \frac{m_\s^2}{\m^2} \right)}_{(4)}\nonumber\\
&& \hspace{4cm}+ \underbrace{ \frac{m_\m^2}{16\p^2 m_\s^2} \left(-1 +\frac{\p^2}{3}\right)}_{(5)}\ ,\nonumber
\end{eqnarray}
which may be expressed as
\begin{eqnarray}
\label{sixteena}
a&=&\underbrace{m_\m^2 \Pi_{\rm EFT}(0) \left(- \frac{1}{\epsilon}-\log \frac{m_\m^2}{\m^2} \right)}_{(1)} \\
&&\hspace{1cm}+ \underbrace{m_\m^2\,  C_{\psi_\m^2A^2}(\m) \left(- \frac{1}{\epsilon}-\log \frac{m_\m^2}{\m^2} \right)}_{(2)}\nonumber\\
&& \hspace{2cm}+\underbrace{m_\m^2  \, C_{\psi_\m^2\p^2}(\m)  \left( -\frac{2}{\epsilon}+\gamma_E+\log \frac{m_\p^2}{4\p\m^2}\right)}_{(3)}\nonumber \\
&&\hspace{4cm}+  \underbrace{m_\m^2 \, C_{\psi_\m^2A}(\m)}_{(4)+(5)}\ ,
\end{eqnarray}
where\footnote{In a minimal subtraction scheme, one drops the combinations  $1/\e+ \mathrm{finite\ constant}$ to obtain the renormalized LECs.}
\begin{subequations}
\begin{eqnarray}
\label{Cs}
C_{\psi_\m^2A^2}(\m)&=&-\frac{1}{16\p^2m_\s^2}\left( \frac{2}{\epsilon}-\gamma_E-\log \frac{m_\s^2}{4\p\m^2}+1 \right)\label{Csa}\ ,\\
C_{\psi_\m^2\p^2}(\m)&=& -\frac{1}{16\p^2m_\s^2} \left(\frac{1}{\epsilon}+\log \frac{m_\s^2}{\m^2} \right)\label{Csb}\ ,\\
C_{\psi_\m^2A}(\m)&=&-\frac{1}{16\p^2m_\s^2} \left(\frac{2}{\e}-\g_E-\log \frac{m_\s^2}{4\p\m^2}+1\right)\left(\frac{1}{\epsilon}+\log \frac{m_\s^2}{\m^2} \right)\\
&&+ \frac{1}{16\p^2 m_\s^2}\left(-1 +\frac{\p^2}{3}\right)\label{Csc}\ ,\nonumber
\end{eqnarray}
Note the presence of terms proportional to $1/\e^2$, as expected at two loops.
where the T4-type LEC $C_{\m^2A^2}(\m)$ already appeared in Eq.~(\ref{T4CT}), and
we encounter the new counter terms
\begin{eqnarray}
\label{T3T5CT}
&T3&:\qquad \frac{\,C_{\psi_\m^2\p^2}(\m)}{4}\, \psi_\m^2\p^2\ ,\\
&T5&:\qquad \frac{C_{\psi_\m^2 A}(\m)}{2}\, \psi_\m^2 A\ .\nonumber
\end{eqnarray}
\end{subequations}
$a^{(1)}$ corresponds to the diagram T2 in Fig.~\ref{Sample2}.
This diagram has two subdivergences, each needing a counter term, one for the pion
loop and one for the muon-photon loop, leading to counter term vertices $\psi_\m^2A^2$ and $\psi_\m^2\p^2$, respectively.  The first vertex, $\psi_\m^2A^2$, is obtained by contracting the
pion loop to a point, and was already encountered in the EFT calculation of $\P(q^2)$;
it corresponds to diagram T4, and leads to $a^{(2)}$.
The second vertex, $\psi_\m^2\p^2$, is new and is obtained by contracting the muon-photon
loop to a point; note that this counter term involves both the pions of the ``strong
interactions'' and the muons of ``QED.''    This new counter term corresponds to diagram
T3, and leads to $a^{(3)}$.
An overall two-loop divergence corresponding to the $1/\e^2$ pole requires a new counter term proportional
to $\psi_\m^2A$. It corresponds to diagram T5 and is represented by the contribution $a^{(4)}$; this counter term can also
have (and does have, in the UV completion provided by our model) a finite part, $a^{(5)}$.
This latter counter term is to be compared to the Pauli term in the case of $a_\m^{\rm HVP}$.  The appearance of $1/\e^2$ poles is consistent with Eq.~(\ref{one}) being a two-loop integral.
Again, the specific result for $a^{(5)}$ is only known because in this example we know the
underlying theory; if we only had access to the EFT, the factor $-1+\p^2/3$ would be replaced by an unknown,
finite constant.   Similar calculations could be carried out to higher orders in $m_\p^2/m_\s^2$ and
$m_\m^2/m_\s^2$.

\subsection{\label{finvol} The toy model in finite volume}
We now turn to the analysis of the interplay between FV effects and UV divergences,
our primary reason for introducing the toy model~(\ref{modelL}) in the first place.   We begin with
calculating FV effects in the model itself, and then calculate and compare them with
the calculation in the EFT developed in the previous subsection.

It is convenient to use the Schwinger parametrization for $\P(q^2)$ given by
\begin{equation}
\label{seventeen}
\Pi(q^2)=\int \frac{d^4p}{(2\p)^4}\int_0^\infty d\alpha_1 \ \alpha_1\int_0^\infty d\alpha_2\ e^{-\alpha_2 (p^2+m_\s^2)}\ e^{-\alpha_1 \left((p-q)^2+m_\p^2\right)}\ .
\end{equation}
Making the change of variables
\begin{equation}
\label{eighteen}
\hat{x}=\frac{\alpha_1}{\alpha_1+\alpha_2}\ ,\quad x=\alpha_1+\alpha_2\quad \longleftrightarrow \quad \alpha_1=x\, \hat{x}\ ,\quad \alpha_2=x(1-\hat{x})\ ,
\end{equation}
one can rewrite $\P(q^2)$ in a finite spatial volume with linear dimension $L$ and periodic boundary conditions as:
\begin{eqnarray}
\label{nineteen}
\Pi_{\rm FV}(q_4^2)&=&\sum_{\vec n} \P^{(n)}(q_4^2)\\
\P^{(n)}(q_4^2)&=&\int_0^1 d\hat{x}\, \hat{x}\int_0^\infty dx \, x^2 \, e^{-xN^2}\int_{-\infty}^{\infty}\frac{dp}{2\p^2}\, p\,\,\int_{-\infty}^{\infty}\frac{dp_4}{2\p}\, e^{-x(p_4^2+p^2)}\,\frac{e^{i np L}}{2 i nL}\ ,\nonumber
\end{eqnarray}
where
\begin{eqnarray}
\label{nineteenA}
N^2&=&q^2 \hx (1-\hx)+ N_0^2\  ,\\
N_0^2&=&m_\s^2(1-\hx)+m_\p^2\hx \ ,\nonumber
\end{eqnarray}
and we made use of the Poisson resummation~(\ref{Afourteen}).
As we are interested here in the FV contributions, we will take $n>0$.
We took $\vec q=0$, as in Sec.~\ref{exfinvol} and shifted $p_4-xq_4\to p_4$, and
$q^2=q_4^2>0$.   Carrying out the integrals over $p_4$ and $p$ yields
\begin{equation}
\label{twentyone}
\P^{(n)}(q^2)=\frac{1}{16\p^2}\int_0^1 d\hat{x}\  \hat{x}\int_0^\infty dx \, e^{-x\, q^2\hx (1-\hx)}\ e^{-x\, N_0^2}\ e^{-n^2L^2/(4 x)}\ .
\end{equation}
We can also write
\begin{equation}
\label{aFV}
a_{\rm FV}=\sum_{\vec n} a^{(n)}\ ,
\end{equation}
so that, performing the integrals over $q^2$ in Eq.~(\ref{one}) and $x$
in Eq.~(\ref{twentyone}), we find
\begin{equation}
\label{an}
a^{(n)}=-\frac{m_\m^2}{16 \p^2}\!\int_0^1 \!\!\!d\hat{x}\,
\frac{\hx}{N_0^2}\left(\!K_0(N_0nL)+N_0nL K_1(N_0nL)\!\left(\!\g_E
+\log{\frac{m_\m^2 n^2L^2\hx(1-\hx)}{2N_0nL}}\right)\!\right) ,
\end{equation}
where $K_\n(z)$ is the modified Bessel function of order $\n$.
Carrying out the integral over $\hx$ (see App.~\ref{integrals} for details), we find for our
final result:
\begin{equation}
\label{twentyfour}
a^{(n)}=m_\m^2\underbrace{\frac{1}{8\p^2}\,\frac{1}{m_\s^2}\,K_0(m_\p nL)}_{\P^{(n)}_{\rm EFT}(0)}\,\log\frac{m_\s^2}{m_\m^2}\ .
\end{equation}
Here we dropped terms $\sim e^{-m_\s nL}$ and terms that are suppressed by additional powers of
$m_\p^2/m_\s^2$ or $1/(m_\s^2 L^2)$.
In Eq.~(\ref{twentyfour}), we already anticipate the result to be derived in Eq.~(\ref{twentysix}) below,
that the prefactor of the logarithm is nothing but $m_\m^2\P^{(n)}_{\rm EFT}(0)$ in finite volume.
 This result conforms with the intuition that FV effects associated with the pions are infrared effects due to the low-energy
degrees of freedom contained in the EFT.   As the $\s$ is not part of the
EFT, its FV effects cannot be obtained from the EFT.

We now turn to the calculation of FV effects in the EFT version of our toy model,
beginning with $\P_{\rm EFT}(0)$.   Defining $\P^{(n)}_{\rm EFT}$ analogous to
the definition of $D^{(n)}$ in Eq.~(\ref{DFV}), we obtain for $n>0$
\begin{eqnarray}
\P^{(n)}_{\rm EFT}(0)&=&\frac{1}{m_\s^2}\int_{-\infty}^{\infty}\frac{ dp_4}{2\p}\int_{-\infty}^{\infty}\frac{ dp}{2\p^2}\frac{p}{\left(p_4^2+p^2+m_\p^2\right)^2} \, \frac{e^{i p nL}}{2 i nL}\ .\label{twentysix}\\
&=&\frac{1}{8\p^2}\,\frac{1}{m_\s^2}\,K_0(m_\p nL)
\nonumber\\
&= &  \frac{1}{8\p^2}\,\frac{1}{m_\s^2}\, \sqrt{\frac{\p}{2}}\,\frac{e^{-m_\p nL}}{\sqrt{m_\p nL}}\ \left( 1+ \mathcal{O}\left( \frac{1}{m_\p nL} \right)  \right)\ ,\nonumber
\end{eqnarray}
where we evaluated
the integral over $p$ using the residue theorem, and then the integral over $p_4$ to yield
the modified Bessel function.

There is another instructive way to obtain the result in Eq.~(\ref{twentysix}). This is by noting that the case $n>0$ is equivalent to putting the particle propagating in the loop on shell. Going to Minkowski space, and after using Eq.~(\ref{Afourteen}) again, one has
\begin{equation}
\P^{(n)}_{\rm EFT}(0)=\left(\frac{-i}{m_\s^2}\right) \int_{-\infty}^{\infty} \frac{dp_0}{2\p}\int_{-\infty}^\infty\frac{dp}{2\p^2}\frac{p}{(p_0^2-p^2-m_\p^2+ i \e)^2}\ \frac{e^{in p L}}{2in L}\label{twentysixb}\ .
\end{equation}
Putting the pion in the loop on shell amounts to the following replacement
\begin{eqnarray}
\frac{1}{(p_0^2-p^2-m_\p^2+ i \e)^2}&=&\frac{1}{2 p}\ \frac{d}{dp}\left(\frac{1}{p_0^2-p^2-m_\p^2+ i \e}\right)\nonumber \\
&\rightarrow& -\frac{i\p}{2 p}\ \ \frac{d}{dp}\,\delta(-p_0^2+p^2+m_\p^2) \label{twentysixc}\ ,
\end{eqnarray}
so that Eq.~(\ref{twentysixb}) becomes
\begin{eqnarray}
\P^{(n)}_{\rm EFT}(0)&=&\frac{-1}{m_\s^2} \left(\frac{1}{8\p^2}\right)\int_{-\infty}^{\infty} dp_0 \int_{-\infty}^{\infty} dp \,\frac{d}{dp}\left[\delta(p_0^2-p^2-m_\p^2)\right] \ \frac{e^{i n p L}}{2 i n L}\nonumber \\
&=&\frac{1}{m_\s^2}\frac{1}{8\p^2}\,\int_{m_\p^2}^\infty \frac{dp_0}{\sqrt{p_0^2-m_\p^2}}\cos\left(n L\sqrt{p_0^2-m_\p^2}\right)\  .  \label{twentysixd}
\end{eqnarray}
A change of variable $y=\sqrt{p_0^2-m_\p^2}$ finally yields
\begin{equation}
\P^{(n)}_{\rm EFT}(0)= \frac{1}{m_\s^2}  \,\frac{1}{16\p^2}\,\int_{-\infty}^\infty \frac{dy}{\sqrt{y^2+m_\p^2}}\ e^{i n L y}\ ,\label{twentysixe}
\end{equation}
which equals Eq.~(\ref{twentysix}).  We note that the
result for $\P^{(n)}_{\rm EFT}(0)$ for $n>0$ is finite; the counter term  $\P_{\rm CT}(0)$ was only needed
in the infinite-volume theory.   It does not involve a pion loop, so there is no contribution to
the FV part of $\P(0)$ in the EFT.   In other words,
$\P_{\rm CT}(0)$ renormalizes $\P_{\rm EFT}(0)$ in
infinite volume, while FV corrections to $\P_{\rm EFT}(0)$ are finite.

The implication of this is that we expect the FV part of $\P_{\rm CT}(0)$ to vanish.
Since we calculated $\P^{(n)}_{\rm EFT}(0)$ to order $1/m_\s^2$ we thus expect that
if we calculate $\P^{(n)}_{\rm CT}(0)$ for $n>0$ in the same way that we calculated
$\P_{\rm CT}(0)$ from the model in Eq.~(\ref{PiCT}) it will turn out not to have a term of
order $1/m_\s^2$, implying that the FV corrections to Eq.~(\ref{PiCT}) vanish.   This is what
we will demonstrate next.

Replacing the $\s$ propagator in Eq.~(\ref{two}) by the second term of Eq.~(\ref{split}),
and again going through the steps to arrive at an expression for the FV correction term
$\P^{(n)}_{\rm CT}(0)$ for $n>0$, we arrive at
\begin{equation}
\P^{(n)}_{\rm CT}(0)=
- \frac{1}{m_\s^2} \int_{-\infty}^{\infty} \frac{dp_4}{2\p}\int_{-\infty}^{\infty} \frac{dp}{2\p^2}\frac{p}{\left(p_4^2+p^2+m_\p^2\right)^2}\left[ \frac{p_4^2+p^2}{p_4^2+p^2+m_\s^2}\right] \frac{e^{i p nL}}{2inL}\  . \label{thirtytwoc}
\end{equation}
In addition to the double poles at $p=\pm i\sqrt{p_4^2+m_\p^2}$ also present in
Eq.~(\ref{twentysix}), the integrand now also has simple poles at $p=\pm i\sqrt{p_4^2+m_\s^2}$.
The pole at $p=i \sqrt{p_4^2+m_\s^2}$ gives a contribution $\sim e^{-L \sqrt{p_4^2+m_\s^2}}$ which, however, is negligible when $m_\s\gg m_\p$, in comparison to the contribution $\sim e^{-L \sqrt{p_4^2+m_\p^2}}$ from the pole at  $p=i \sqrt{p_4^2+m_\p^2}$, and may thus
be discarded.\footnote{This reflects the fact that  light particles dominate the FV effects in the EFT for our toy model.}
Furthermore, we can immediately see why the remaining part of $\P^{(n)}_{\rm CT}(0)$
is suppressed  relative to $\P^{(n)}_{\rm EFT}(0)$ by an extra factor $m_\p^2/m_\s^2$.   The suppression by this factor follows from the presence of the term in
square brackets in Eq.~(\ref{thirtytwoc}) and dimensional analysis.   The factor $e^{-L\sqrt{p_4^2+m_\p^2}}$ present after integration over $p$ makes the integral over
$p_4$ sufficiently convergent that the rest of the integrand can be expanded in
inverse powers of $m_\s^2$.   We conclude that, to the order we are working,
indeed, no counter-term FV correction is produced by the underlying model.
We note that, in this argument, we always assume that $m_\p L\gg 1$, \ie, that we
consider FV effects in the $p$-regime.

To summarize, to leading order in an expansion in $m_\p^2/m_\s^2$, we conclude
that, for $n>0$,
\begin{equation}
\label{summ}
\P^{(n)}_{\rm CT}(0)=0\qquad \Rightarrow\qquad \P^{(n)}(0)=\P^{(n)}_{\rm EFT}(0)\ .
\end{equation}
No counter term is needed for the FV corrections to $\P(0)$ calculated using the EFT.

Now, let us return to our ``anomaly,'' $a$ defined in Eq.~(\ref{one}).
First, substituting Eq.~(\ref{twentysix}) into Eq.~(\ref{one}) one finds,
for the finite-volume part of $a$ in the EFT,
\begin{eqnarray}
\label{twentyseven}
a^{(n)}_{\rm EFT}&=&\int_0^\infty dq^2\left(\frac{q^2}{\m^2}\right)^\e \, \frac{m_\m^2}{q^2+m_\m^2}\  \P^{(n)}_{\rm EFT}(0)\nonumber \\
&=& m_\m^2 \P^{(n)}_{\rm EFT}(0)  \left( - \frac{1}{\epsilon} - \log m_\m^2\right)\nonumber\\
&=& m_\m^2 \P^{(n)}(0) \left( - \frac{1}{\epsilon} - \log \frac{m_\m^2}{\m^2}\right)\  .
\end{eqnarray}
This corresponds to the FV part of $a_1$ defined in Eq.~(\ref{twelve}):  it is the FV contribution from
the pion loop in diagram T2 in Fig.~\ref{Sample2}.   Clearly, a counter-term contribution will
be needed to make this FV contribution finite.   The counter term needed here is the same
infinite-volume counter term used in diagram T3 of Fig.~\ref{Sample2},
with the pion loop in that diagram producing the FV part.   Diagrams T4 and T5 do not
contribute to the pion-induced FV corrections (T4 would only yield FV contributions due
to a muon wrapping around the world).
We emphasize how it is consistent to keep the $q^2$ integral in Eq.~(\ref{twelve})  in infinite volume, even though $\Pi(0)$ itself is replaced by its FV part.
In other words, we can treat the ``strong-interaction" (\ie, $\p,\s$) physics in finite volume, while keeping the ``electromagnetic''
(\ie, $A,\m$) physics in infinite volume.

Finally, as before, we need to consider $a_2$ given by ({\it cf.} Eq.~(\ref{twelve}))
\begin{equation}
\label{thirtyseven}
a_2=m_\m^2 \int_0^\infty \frac{dq^2}{q^2+\underbrace{m_\m^2}_{\mathrm{neglect}}}\ \left(\frac{q^2}{\m^2}\right)^\epsilon \left\{   \Pi(q^2)-\Pi(0)\right\}\ ,
\end{equation}
but now replacing the pion-physics part, $\P(Q^2)-\P(0)$, by its FV correction.
As before, the $m_\m$ in the denominator can be neglected because $\Pi(q^2)-\Pi(0)\sim q^2$ as $q^2\to 0$. The term proportional to $\Pi(0)$ in Eq.~(\ref{thirtyseven}) then vanishes in dimensional regularization, so the net result becomes
\begin{equation}
\label{thirtyeight}
a^{(n)}_2= m_\m^2 \int_0^\infty \frac{dq^2}{q^2}\,\left(\frac{q^2}{\m^2}\right)^\e\, \Pi^{(n)}(q^2)\ .
\end{equation}
Inserting Eq.~(\ref{twentyone}) and using the method of App.~\ref{integrals} that was used to
evaluate Eq.~(\ref{an}),\footnote{For the $1/\e$ part, all that is needed is $K_1(y)=-dK_0(y)/dy$.}
we obtain
\begin{equation}
\label{thirtynine}
a^{(n)}_2
=m_\m^2 \P^{(n)}(0)  \left(\frac{1}{\e}+ \log  \frac{m_\s^2}{\m^2}  \right)\ .
\end{equation}
An important remark here is that, even though the end result~(\ref{thirtynine}) is proportional 
to $\P^{(n)}(0)$, the $q^2$ dependence of $\P^{(n)}(q^2)$ in the integrand in Eq.~(\ref{thirtyeight})
is crucial for obtaining this result.
The factor $1/\e+\log(m_\s^2/\m^2)$ is the same as that appearing in $a^{(3)}$ in infinite volume, on the third and fourth line of Eq.~(\ref{sixteen}).
Thus, adding Eqs.~(\ref{twentyseven}) to~(\ref{thirtynine}), one arrives at
\begin{equation}
\label{forty}
a^{(n)}=a^{(n)}_1+a^{(n)}_2= m_\m^2 \Pi^{(n)}(0)  \log\frac{m_\s^2}{m_\m^2}\ ,
\end{equation}
where $\P^{(n)}(0)= \P^{(n)}_{\rm EFT}(0)$ and $\P^{(n)}_{\rm EFT}(0)$ is given in Eq.~(\ref{twentysix}). This is equal to the complete result~(\ref{twentyfour}) calculated in the underlying toy model. In summary, the volume dependence is contained in the contribution from the EFT, \ie, $\P^{(n)}(0)$, but the infinite-volume
counter term with coefficient $C_{\psi_\m^2\p^2}(\m)$ given in Eq.~(\ref{Csb}) is needed to render the FV
contribution finite.  The contributions
$a^{(4)}$ and $a^{(5)}$ in Eq.~(\ref{sixteen}) do not contribute pion-induced FV effects, as
the corresponding diagrams T4 and T5 in Fig.~\ref{Sample2} do not contain pion loops.

\section{\label{conclusion} Conclusion}
%%%%%%%%%%%%%%%%%%%%%%%%%%%
In this paper, we considered the effective field theory approach to the hadronic
vacuum polarization contribution to the muon anomalous magnetic moment, $a_\m^{\rm HVP}$.
Our primary aim was a better understanding of how the evaluation of finite-volume
effects works in an EFT framework, but this led us to a discussion of the counter-term
structure needed for a complete EFT representation of $a_\m^{\rm HVP}$\ in infinite volume.
The specific motivation
for our interest in finite-volume effects is to contribute to a deeper understanding of the
systematic error caused by these effects in lattice computations of $a_\m^{\rm HVP}$.

Finite-volume effects in lattice computations of $a_\m^{\rm HVP}$\ are dominated by pions, and thus the natural EFT
candidate is chiral perturbation theory.   As we showed, for a complete low-energy description
of $a_\m^{\rm HVP}$, also muons and photons have to be included in the EFT.   In particular,
since $a_\m^{\rm HVP}$\ also contains a loop with muons and photons, counter terms in our EFT
can contain not only pion fields, but also muon and photon fields.   Just like counter
terms are needed to regulate the UV behavior of pion loops in ChPT, additional counter
terms are also needed to regulate the UV behavior of the muon-photon loop.   These
counter terms lead to the presence of new low-energy constants in the EFT, in
addition to those already present in ChPT.   The values of these LECs can, of course,
only be  fixed by matching with the underlying UV-complete theory, QCD plus QED.

Once these LECs are taken into account,
ChPT, augmented with muon and photon fields, gives a complete representation of
$a_\m^{\rm HVP}$.   In addition, finite-volume effects at any given order
are completely fixed in terms of the LECs
that appear to one order less in the EFT.    In this sense, finite-volume effects can
be predicted, at any given order, in terms of the LECs of the infinite-volume EFT.
This is not surprising, as $a_\m^{\rm HVP}$\ is nothing else than (the projection to zero
external photon momentum and onto the Pauli spin structure of) a correlation function
in QCD plus QED.

In the case of $a_\m^{\rm HVP}$, we explained that the need for new counter terms, not
present in ChPT, the EFT for pions alone, arises for the first time in the EFT
expansion at N$^3$LO, and that their
role in the study of finite-volume effects starts at N$^4$LO.   We constructed the
new counter terms at the lowest order at which they appear, and showed that
indeed dimensional arguments imply that they become relevant at N$^3$LO.
While the construction of these counter terms is relatively straightforward,
it would require three- and four-loop calculations to demonstrate how it all
works in the case of $a_\m^{\rm HVP}$.   Therefore, instead, we demonstrated our
observations by working out a toy model in which the
effects appear already at two loops, which is the minimum number.    While it is unlikely that the full N$^3$LO
analysis will ever be worked out in practice, it is important to establish the
validity of the EFT framework for the study of $a_\m^{\rm HVP}$ in order to be assured that
even the NNLO analysis of finite-volume effects carried out in Ref.~\cite{us2} has a
solid EFT basis.   We believe that our discussion in this paper illustrates why
indeed this is the case.
We expect the same separation between the UV 
physics represented by the counter terms and IR physics of finite-volume effects 
will also take place in the hadronic light-by-light contribution when the QED part 
is taken in infinite volume \cite{HLbL}. 

\vspace{2ex}
%\newpage
\noindent {\bf Acknowledgments}
\vspace{2ex}
%%%%%%%%%%%%%%%%%%%%%%%%%%%

We thank Max Hansen for discussions.  TB’s and MG’s work is supported by the U.S. Department of Energy, Office of Science, Office of High Energy Physics, under Award Numbers DE-SC0010339, DE-SC0013682, respectively. S.P. is supported by CICYTFEDER-FPA2017-86989-P and by Grant No. 2017 SGR 1069.

\appendix
\section{\label{timemom} Counter terms in the time-momentum representation}
In this appendix, we revisit the time-momentum representation for the HVP \cite{BM}.
In lattice computations, $a_\m^{\rm HVP}$\ is often obtained from
\begin{equation}
\label{Ct}
C(t)=\frac{1}{3}\sum_i\int d^3x\svev{j_i(0,0)j_i(\vec x,t)}\ ,
\end{equation}
where $j_\m$ is the hadronic part of the electromagnetic current.   We can, of course,
express $C(t)$ in terms of the HVP $\P(q^2)$:
\begin{eqnarray}
\label{CHVP}
C(t)
&=&\frac{1}{3}\sum_i\int d^3x\int\frac{d^4q}{(2\p)^4}\,e^{iqx}\left(\d_{ii}q^2-q_iq_i\right)\P(q^2)\\
&=&\int\frac{dq_4}{2\p}\,e^{iq_4 t}\,q_4^2\P(q_4^2)\nonumber\\
&=&-\int\frac{dq_4}{2\p}\,e^{iq_4 t}\,q_4^2\hat\P(q_4^2)-\P(0)\d''(t)\ ,\nonumber
\end{eqnarray}
where
\begin{equation}
\label{Pihat}
\hat\P(q^2)=\P(0)-\P(q^2)\ .
\end{equation}
Equation~(\ref{CHVP}) contains the divergent counter term $-\P(0)\d''(t)$, and we note that,
in general, counter terms in $C(t)$ live at $t=0$,
and thus appear in the form of the Dirac delta function $\d(t)$ and its derivatives.
In particular, in ChPT, higher derivatives than $\d''(t)$ can appear, corresponding to the
fact that counter terms are polynomials in $q^2$ in the momentum representation.

Inverting Eq.~(\ref{CHVP}), we find
\begin{equation}
\label{qsqPi}
q_4^2\P(q_4^2)=\int_{-\infty}^\infty dt\left(e^{-iq_4 t}-1\right)C(t)
=-4\int_0^\infty dt\,\sin^2(q_4 t/2)C(t)\ ,
\end{equation}
where we inserted the $-1$ to enforce that the LHS vanishes for $q_4=0$, and we
used that $C(t)$ is even in $t$ in the second step.   It follows that
\begin{equation}
\label{PihatC}
\hat\P(q_4^2)=\int_0^\infty dt\left(\frac{4\sin^2(q_4 t/2)}{q_4^2}-t^2\right)C(t)\ ,
\end{equation}
and thus
\begin{eqnarray}
a^{\rm HVP}_\m&=&\int_0^\infty dq^2 f(q^2) \hat\P(q^2)\label{5one}\\
&=&  \int_0^\infty dq^2 f(q^2) \int_0^\infty dt\left(\frac{4\sin^2(q\, t/2)}{q^2}-t^2\right)C(t)\ .\nonumber
\end{eqnarray}
In Ref.~\cite{us2} we made the incorrect observation that at NNLO $a_\m^{\rm HVP}$\ is UV divergent in ChPT,
because at NNLO $C(t)\sim 1/t^5$ for $t\to 0$, which makes the integral over $t$ in Eq.~(\ref{5one}) divergent,
as the weight function in the integral over $t$ behaves like $t^4$ for $t\to 0$.   If true, this would
be a paradox, because, as we have seen in Sec.~\ref{intro}, the expression for $a_\m^{\rm HVP}$\ in
terms of $\hat\P(q^2)$ is UV finite at NNLO in ChPT.   The problem, in fact, has nothing to do
with $a_\m^{\rm HVP}$, as can be seen from Eq.~(\ref{PihatC}).   While $\hat\P(q^2)$ is finite by construction,
the integral on the right-hand side of this equation is UV divergent if indeed $C(t)\sim 1/t^5$
for small $t$ at NNLO.

As we will show here, this apparent paradox arises if one is not careful with the treatment of
counter terms in the time-momentum representation.
The correct expression connecting  $\widehat{\Pi}(q^2)$ and $C(t)$ in ChPT is
\begin{equation}
\label{right}
\hat\P(q^2)=\lim_{\e\to 0}\int_{0^-}^\infty dt\left(\frac{4\sin^2(q t/2)}{w^2}-t^2\right)\left(C_{\rm ChPT}(t;\e,\m)- \frac{1}{2}\,\delta^{IV}(t) \ C(\e,\m)\right)\ ,
\end{equation}
where $\delta^{IV}(t) $ is the 4th derivative of the Dirac delta and $C(\e,\m)$ is a counter term directly related to the LECs of ChPT.  As usual, in order to define a counter term, a regulator
needs to be introduced, and the parameter $\e$ represents this regulator, as we will see below.
The rest of this appendix will prove this result.   While we keep our discussion concrete by using a simple example to make the argument, the end result carries over to the case of ChPT.

In the following, it will be useful to express $C(t)$ in terms of the spectral function $\r(s)$.
From the subtracted dispersion relation
\begin{equation}
\label{disp}
\hat\P(q^2)=q^2\int_0^\infty ds\,\frac{\r(s)}{s(s+q^2)}\ ,
\end{equation}
it follows from Eq.~(\ref{CHVP}) that
\begin{eqnarray}
\label{Ctagain}
C(t)&=&- \int_0^\infty ds\, \r(s) \int_{-\infty}^\infty \frac{dq_4}{2\p}\, e^{iq_4 t} \frac{q_4^4}{s(s+q_4^2)}-\P(0)\d''(t)\\
&=& - \int_0^\infty ds\, \r(s)\left(-\d(t)-\frac{1}{s}\,\d''(t)+\half\sqrt{s}\,e^{-\sqrt{s}|t|}\right)-\P(0)\d''(t)\nonumber\\
&=&- \int_0^\infty ds\, \r(s)\left(-\d(t)+\half\sqrt{s}\,e^{-\sqrt{s}|t|}\right)\ ,\nonumber
\end{eqnarray}
where in the last step we used the unsubtracted dispersion relation.

To illustrate the claim and understand what is going on, let us 
consider the simple example of a spectral function given by
\begin{equation}\label{ex0}
  \rho(s)=\left( 1+ \frac{s}{s+M^2}\right) \theta(s-m^2)\ ,
\end{equation}
where $M$ is like the $\r$-meson mass, for example, and we are interested in a ``ChPT'' result where $M^2\gg m^2\sim q^2 \equiv m_{L}^2$ ($L$ for light). With this spectral function, one may calculate the once-subtracted vacuum polarization, $\hat{\Pi}(Q^2)$ as
\begin{eqnarray}
  \hat{\P}(q^2)&=&q^2 \int_{m^2}^{\infty}\frac{ds}{s(s+q^2)}\, \rho(s) \label{ex2a}\\
  &=&\log\frac{m^2+q^2}{m^2}+\frac{q^2}{M^2-q^2}\log \frac{m^2+M^2}{m^2+q^2}\nonumber\\
  &\approx & \log\frac{m^2+q^2}{m^2}+\frac{q^2}{M^2}\log \frac{M^2}{m^2+q^2}+ \mathcal{O}\left(\frac{m_{L}^4}{M^4}\right)\ . \nonumber
\end{eqnarray}
In this world, the identification of the counter term can be made by splitting the spectral function as
\begin{equation}
\label{ex0a}
\rho(s)=\underbrace{1+\frac{s}{M^2}}_{\rho_{\rm ChPT}(s) }\ \underbrace{- \frac{s^2}{M^2(s+M^2)}}_{\Delta\rho(s)}
\end{equation}
in terms of the ``ChPT'' spectral function, $\r_{\rm ChPT}(s)$, and the ``UV completion," $\D\r(s)$.  With this split, also the dispersion relation~(\ref{ex2a}) splits into two parts:
\begin{eqnarray}
\hat{\Pi}(q^2)&=& q^2 \int_{m^2}^\infty \frac{ds}{s(s+q^2)}\left(  1+ \frac{s}{M^2}\right) \left( \frac{s}{\m^2}\right)^\e \label{ex3a}\\
&&\hspace{1cm} +\ q^2 \int_{\underbrace{m^2}_{0}}^\infty \frac{ds}{s(s+\underbrace{q^2}_{0})}\left(  - \frac{s^2}{M^2(s+M^2)}\right)  \left( \frac{s}{\m^2}\right)^\e \nonumber\\
&\equiv & \hat{\Pi}_{\rm ChPT}(q^2,\e,\m) + q^2\, C(\e,\m) \nonumber\ ,
\end{eqnarray}
where
\begin{equation}
\hat{\P}_{\rm ChPT}(q^2,\e,\m)=q^2 \int_{m^2}^\infty \frac{ds}{s(s+q^2)}\left(  1+ \frac{s}{M^2}\right) \left( \frac{s}{\m^2}\right)^\e \label{ex3d}
\end{equation}
and
\begin{equation}
C(\e,\m)=\int_{\underbrace{m^2}_{0}}^\infty \frac{ds}{s(s+\underbrace{q^2}_{0})}\left(  - \frac{s^2}{M^2(s+M^2)}\right)  \left( \frac{s}{\m^2}\right)^\e \quad .\label{ex3e}
\end{equation}
We have added a regulating factor $(q^2/\m^2)^\e$ because both integrals are now separately UV divergent. In the second integral, Eq.~(\ref{ex3e}), we may take the limit $q^2,m^2\to 0$ as it is IR convergent thanks to the subtraction $\Delta\rho(s)$, making $C(\e,\m)$ depend only on the UV scale $M$. Both integrals may be evaluated to give:
\begin{equation}
\hat{\Pi}(q^2)=\left(1-\frac{q^2}{M^2}\right) \log\left(1+\frac{q^2}{m^2}\right) - \frac{q^2}{M^2}\left( \frac{1}{\e}+ \log\frac{m^2}{\m^2}\right) +q^2\, C(\e,\m)\ ,\label{ex4}
\end{equation}
and
\begin{equation}
C(\e,\m)=\frac{1}{M^2}\left(\frac{1}{\e}+\log \frac{M^2}{\m^2}\right)\ .\label{ex5}
\end{equation}
$C(\e,\m)$ is the necessary counter term,  and encodes the UV completion of the theory. One can check that the combined result~(\ref{ex4}) agrees with the last line of Eq.~(\ref{ex2a}), as it should.

Let us now discuss consider Eq.~(\ref{right}), using our example. Using Eqs.~(\ref{Ctagain}) and~(\ref{ex0a}):
\begin{eqnarray}
C_{\rm ChPT}(t)&=&-\int_{m^2}^\infty ds\ \rho_{\rm ChPT}(s) \left(\frac{\sqrt{s}}{2}\ e^{-t\sqrt{s}}-\delta(t)\right)\quad (t> 0^-)\label{ex5a} \\
&&\hspace{-3cm}=  \frac{e^{-m t}}{t^5}\left[t^2\left(-2-m t \left( 2+ m t \right)\right)+ \frac{-24-m t\left(24+m t \left(12+ m t \left(4+ m t\right)\right)\right)}{M^2}  \right]\nonumber\\
&&\hspace{-3cm}\approx-\frac{2}{t^3} - \frac{24}{M^2 t^5}\qquad (t\to 0^+)\ . \nonumber
\end{eqnarray}
The term proportional to $\delta(t)$ will not contribute to $\widehat{\Pi}(q^2)$ in Eq.~(\ref{right}), since the kernel $\left(\frac{4\sin^2(w t/2)}{w^2}-t^2\right)\sim t^4$, and we will thus neglect this
term in what follows. The result in Eq.~(\ref{ex5a}) shows that our example~(\ref{ex0}) reproduces the $1/t^5$ behavior  for $t\to 0$ of ChPT at NNLO in the real world and, consequently, it also makes  $\widehat{\Pi}(q^2)$ in Eq.~(\ref{PihatC}) diverge. In view of Eq.~(\ref{ex2a}), this result is clearly incorrect.

First, using for instance Eq.~(\ref{ex3d}), we need to regulate this divergence.\footnote{Alternatively one could also introduce a $t^\eta$ for $\eta\to 0$ in the integrand of Eq.~(\ref{ex5a})  as a regulator. The difference will be a redefinition of the counter term $C(\e,\m)$ in Eq.~(\ref{ex5}) but the final result for $\hat{\P}(q^2)$ will be the same.} The result of Eq.~(\ref{ex3d}) may be expressed as:
\begin{equation}
\hat{\Pi}_{\rm ChPT}(q^2,\e,\m)=\int_0^\infty dt \left(\frac{4\sin^2(q t/2)}{q^2}-t^2\right)\ C_{\rm ChPT}(t;\e,\m)\ ,\label{ex6} \\
\end{equation}
with
\begin{equation}
\label{CChPTdef}
C_{\rm ChPT}(t;\e,\m)=
-\int_{m^2}^\infty ds\ \left(1+\frac{s}{M^2}\right)  \,\frac{\sqrt{s}}{2}\  e^{-t\sqrt{s}}\ \left(\frac{s}{\m^2}\right)^\e \ .
\end{equation}
Then, using the identity
\begin{equation}
q^2=\int_{0^-}^\infty dt \left(\frac{4\sin^2(q t/2)}{q^2}-t^2\right) \left(-\frac{1}{2}\ \delta^{IV}(t) \right)\ ,\label{ex7}
\end{equation}
we can express the result in Eq.~(\ref{ex4}) as
\begin{equation}
\hat\P(q^2)=\int_{0^-}^\infty dt\left(\frac{4\sin^2(q t/2)}{q^2}-t^2\right)\left(C_{\rm ChPT}(t;\e,\m)- \frac{1}{2}\,\delta^{IV}(t) \ C(\e,\m)\right)\label{ex8}\ ,
 \end{equation}
which is precisely Eq.~(\ref{right}), as promised.
To verify this, it is easiest to do the $t$ integral first, then express the integral over
$s$ in terms of the Beta-function, and finally, to use the identities ($x>0$ and $\e<0$)
\begin{eqnarray}
\label{Beta}
(-1)^{1+\e}B(-x,-\e,0)&=&\frac{1}{\e}+\log\frac{1+x}{x}+\co(\e)\ ,\\
(-1)^{1+\e}B(-x,1-\e,0)&=&\log(1+x)+\co(\e)\ .\nonumber
\end{eqnarray}

Our example generalizes to ChPT, because it shares the $1/t^5$ behavior at NNLO.
In order to properly treat the divergence arising in Eq.~(\ref{right}), $C_{\rm ChPT}(t)$
needs to be properly regulated, and thus a counter term needs to be introduced
to define the integral when the
regulator is removed.   In momentum space, the counter term is proportional to
$q^2$, as follows from Eq.~(\ref{ex7}), and it corresponds to the LEC $c_{56}$ in two-flavor
ChPT \cite{BCE}.

\section{\label{integrals} Integrals}
In this appendix, we provide the details of some calculations in the main text.
We begin with a derivation of the explicit form of the third line of Eq.~(\ref{tent}).
Starting from Eq.~(\ref{seventb}), the additional term containing $m_\p^2$ can be written as
\begin{equation}
\label{Bsixteen}
\delta a_{\rm CT}=\frac{m^4}{f_\p^4}\int_0^1 dx \int_0^\infty \frac{dq^2}{q^2+M^2}\,(q^2)^\e\ \log \frac{q^2 x (1-x) + m_\p^2}{q^2 x (1-x)}\ ,
\end{equation}
which is finite so that the limit $\e\to 0$ may be taken. One notices that the ``heavy propagator'' admits a Mellin--Barnes representation
\begin{eqnarray}
\label{Bseventeen}
\frac{1}{q^2+M^2}&=& \frac{1}{M^2}\sum_{n=0}^\infty(-1)^n\left(\frac{q^2}{M^2}\right)^n\\
&=&
 \frac{1}{M^2}\,\frac{1}{2\p i}\int_\gamma ds \left(\frac{q^2}{M^2}\right)^{-s} \p \csc\left(\p s\right)\ , \quad 0<\mbox{Re}\ s<1\ ,\nonumber
\end{eqnarray}
where $\g$ is a line parallel to the imaginary $s$-axis with $0<\mbox{Re}\ s<1$.
(One can see this by using
\begin{equation}
\label{csc}
\p\csc(\p s)=\sum_{n=-\infty}^\infty\frac{(-1)^n}{n+s}\ ,
\end{equation}
and then closing the contour in the left half plane counter clockwise.)
Then,
\begin{eqnarray}
\delta a_{\rm CT}&=&\frac{m^4}{f_\p^4}\ \frac{1}{M^2}\ \frac{1}{2\p i}\int_\gamma ds \, \p \csc\left(\p s\right)\int_0^1 dx \int_0^{\infty} dq^2\left(\frac{q^2}{M^2}\right)^{-s}\log\frac{q^2 x (1-x)+m^2}{q^2 x (1-x)}\nonumber \\
&=&\frac{m^4}{f_\p^4}\ \frac{1}{2\p i}\int_\gamma ds \left(\frac{m^2}{M^2}\right)^{1-s} \Phi(s) \quad ,\label{Beighteen}
\end{eqnarray}
where
\begin{equation}
\label{Bnineteen}
\Phi(s)= \frac{\p^2 \csc^2\left(\p s\right) \Gamma(s)^2}{(1-s) \Gamma(2s)}\asymp \frac{2}{s^3}+\frac{2}{s^2}+ \frac{6+\p^2}{3}\frac{1}{s}+\co(s^0)\ , \quad s\to 0 \ .
\end{equation}
The expansion after the symbol ``$\asymp$" is the singular expansion of the function $\Phi(s)$. According to the Converse Mapping Theorem \cite{FGD}, the expansion for $\delta a_{\rm CT}$
in powers of $m^2/M^2$ is then
\begin{equation}
\label{Btwenty}
\delta a_{\rm CT}=  \frac{m^4}{f_\p^4}\,\frac{m^2}{M^2}\left(\log^2\frac{m^2}{M^2}-2 \log\frac{m^2}{M^2}+ \frac{6+\p^2}{3}  \right) \ .
\end{equation}
The function $\Phi(s)$ has poles not only at $s=0$, but also at all the negative integers
in the left half plane, but those poles lead to higher powers of $m^2/M^2$.   For example,
near $s=-1$ the singular expansion is
\begin{equation}
\label{Btwentyone}
\Phi(s)= \frac{\p^2 \csc^2\left(\p s\right) \Gamma(s)^2}{(1-s) \Gamma(2s)}\asymp \frac{2}{(s+1)^3}-\frac{2}{(s+1)^2}+ \frac{-15+2\p^2}{6}\frac{1}{s+1}+\co((s+1)^0)\ ,
\end{equation}
which leads to a contribution of order $(m^4/f_\p^4)(m^4/M^4)$ times logarithms to
$\delta a_{\rm CT}$.

Next, we give some details about the derivation of Eq.~(\ref{twentyfour}) from Eq.~(\ref{an}).
For simplicity, we set $n=1$; the dependence on $n$ can be restored in the final
result by replacing $L\to nL$.
A change of variables
\begin{equation}
\label{Atwo}
y^2=\lmssq (1-\hx) + \lmpsq \hx \quad \leftrightarrow \quad \hx=\frac{y^2-\lmssq}{\lmpsq-\lmssq} \quad ,
\end{equation}
allows us to rewrite Eq.~(\ref{twentyfour}) as
\begin{eqnarray}
\label{step1}
a^{(1)}&=&-\frac{m_\m^2}{16\p^2}\frac{2}{(m_\s^2-m_\p^2)^2 L^2}\int_{m_\p L}^{m_\s L}dy\,(m_\s^2L^2-y^2)\\&&\times\left(\frac{K_0(y)}{y}+K_1(y)\left(\g_E+\log\frac{m_\m^2(m_\s^2L^2-y^2)(y^2-m_\p^2L^2)}{2(m_\s^2-m_\p^2)^2L^2y}\right)\right)\ .\nonumber
\end{eqnarray}
Next, we will always drop terms that are manifestly suppressed by $m_\p^2/m_\s^2$ or
$e^{-m_\s L}$ relative to the dominant contribution. 
Therefore, we can replace the upper limit $m_\s L$ of the integral by $\infty$ and we may neglect 
$y^2$ in comparison with  $m_\s^2 L^2$ in the combinations $m_\s^2 L^2-y^2$ appearing in Eq.~(\ref{step1}), because the functions $K_{0,1}(y)$ fall off exponentially like $e^{-y}$ at large $y$ and thus suppress such contributions by a factor $e^{-m_\s L}$.   This simplifies Eq.~(\ref{step1})
to
\begin{equation}
a^{(1)}\approx -\frac{1}{8\p^2}\, \frac{m_\m^2}{m_\s^2} \int_{m_\p L}^\infty \frac{dy}{y}\left[ K_0(y)+y\, K_1(y) \left(\gamma_E + \log \frac{m_\m^2 (y^2-m_\p^2 L^2)}{2 m_\s^2 y}\right)\right]\ ,\label{step2}
\end{equation}
and, using the representation
\begin{equation}
\label{step3}
K_\n(y)=\frac{y^\n}{2}\int_{-\infty}^\infty dz\ \frac{e^{i z}}{(z^2+y^2)^{\n+\frac{1}{2}}}\quad \quad (\n=0,1) \  ,
\end{equation}
one may carry out the $y$ integral and obtain
\begin{eqnarray}
a^{(1)}&\approx&  - \frac{1}{16\p^2}\, \frac{m_\m^2}{m_\s^2} \int_{-\infty}^\infty dz\ \frac{e^{i z}}{(z^2+m_\p^2 L^2)^{\frac{1}{2}}}\left[ \gamma_E + \log \frac{2 m_\m^2 (z^2+m_\p^2 L^2)}{m_\s^2 m_\p L}\right]\nonumber \\
&=& \frac{1}{8\p^2} \frac{m_\m^2}{m_\s^2}\ K_0(m_\p L) \log\frac{m_\s^2}{m_\m^2}\ .
\label{step4}
\end{eqnarray}
Finally, the replacement $L\to n L$ yields the result (\ref{twentyfour}),  as promised.

\end{document}